\documentclass[twocolumn]{svjour3}
\usepackage[colorlinks=true,urlcolor=blue,citecolor=black,linkcolor=black]{hyperref}
\usepackage{color}
\usepackage{url}
\usepackage{wrapfig}
\usepackage{algorithm}
\usepackage{algpseudocode}
\usepackage{graphicx}
\usepackage{enumitem}
\usepackage{epsfig}
\usepackage[dvipsnames]{xcolor}
\usepackage{latexsym, amsmath, amsfonts, amssymb, bbm}
\usepackage{balance}
\usepackage{subcaption}

\usepackage{multirow}

\usepackage{graphicx}
\usepackage{subcaption}

\usepackage{tabularx}
\usepackage{colortbl}
\usepackage{longtable}
\usepackage{booktabs}
\usepackage{array}
\usepackage{threeparttable}
\usepackage[pagewise]{lineno}

\usepackage[most]{tcolorbox}

\usepackage{xparse}
\usepackage{lipsum}
\def\exampletext{Example} 
\NewDocumentEnvironment{testexample}{O{}}
    {
    \colorlet{colexam}{blue!55!black} 
        \newtcolorbox[use counter=testexample]{testexamplebox}{
        empty,
        title={\exampletext: #1},
        attach boxed title to top left,
        minipage boxed title,
        boxed title style={empty,size=minimal,toprule=0pt,top=4pt,left=3mm,overlay={}},
        coltitle=colexam,fonttitle=\bfseries,
        before=\par\medskip\noindent,parbox=false,boxsep=0pt,left=3mm,right=0mm,top=2pt,breakable,pad at break=0mm,
        before upper=\csname @totalleftmargin\endcsname0pt,
        overlay unbroken={\draw[colexam,line width=.5pt] ([xshift=-0pt]title.north west) -- ([xshift=-0pt]frame.south west); },
        overlay first={\draw[colexam,line width=.5pt] ([xshift=-0pt]title.north west) -- ([xshift=-0pt]frame.south west); },
        overlay middle={\draw[colexam,line width=.5pt] ([xshift=-0pt]frame.north west) -- ([xshift=-0pt]frame.south west); },
        overlay last={\draw[colexam,line width=.5pt] ([xshift=-0pt]frame.north west) -- ([xshift=-0pt]frame.south west); },
        }
}

\usepackage{framed}
\definecolor{light-gray}{gray}{0.9}
\usepackage[framemethod=TikZ]{mdframed}
\mdfsetup{skipabove=5pt,skipbelow=3pt}
\usepackage{lipsum}
\mdfdefinestyle{MyFrame}{%
    linecolor=black,
    outerlinewidth=0.15pt,
    roundcorner=3pt,
    innertopmargin=2pt,
    innerbottommargin=2pt,
    innerrightmargin=4pt,
    innerleftmargin=4pt,
    backgroundcolor=light-gray}
    
\mdfdefinestyle{MyFrameSimple}{%
    linecolor=black,
    outerlinewidth=0.15pt,
    roundcorner=3pt,
    innertopmargin=2pt,
    innerbottommargin=2pt,
    innerrightmargin=4pt,
    innerleftmargin=4pt}

\newcommand{\sectopic}[1]{\vspace{0em}\par\noindent{\textit{\bfseries #1}}}

\usepackage{etoolbox}
\newcommand*{\affaddr}[1]{#1} 
\newcommand*{\affmark}[1][*]{\textsuperscript{#1}}

\begin{document}

\title{Improving Requirements Completeness: Automated Assistance through Large Language Models}


\author{
		Dipeeka~Luitel\affmark[1] \and
            Shabnam~Hassani\affmark[1]  \and
            Mehrdad~Sabetzadeh\affmark[1] 
}
\institute{{Dipeeka Luitel, Shabnam Hassani, Mehrdad Sabetzadeh \at \email{Dipeeka.Luitel, S.Hassani, M.Sabetzadeh @uottawa.ca}}      \\
\affaddr{\affmark[1]School of Electrical Engineering and Computer Science, University of Ottawa, Canada}\\ \\
}


\maketitle

\thispagestyle{plain}
\pagestyle{plain}

\section*{Abstract}\label{sec:abstract}

Natural language (NL) is arguably the most prevalent medium for expressing systems and software requirements.
Detecting incompleteness in NL requirements is a major challenge. One approach to identify incompleteness is to compare requirements with external sources. Given the rise of large language models (LLMs), an interesting question arises: Are LLMs useful external sources of knowledge for detecting potential incompleteness in NL requirements? This article explores this question by utilizing BERT. Specifically, we employ BERT's masked language model (MLM) to generate contextualized predictions for filling masked slots in requirements. To simulate incompleteness, we withhold content from the requirements and assess BERT's ability to predict terminology that is present in the withheld content but absent in the disclosed content. BERT can produce multiple predictions per mask. Our first contribution is determining the optimal number of predictions per mask, striking a balance between effectively identifying omissions in requirements and mitigating noise present in the predictions. Our second contribution involves designing a machine learning-based filter to post-process BERT's predictions and further reduce noise. We conduct an empirical evaluation using 40 requirements specifications from the PURE dataset. Our findings indicate that: (1)~BERT's predictions effectively highlight terminology that is missing from requirements, (2)~BERT outperforms simpler baselines in identifying relevant yet missing terminology, and (3)~our filter reduces noise in the predictions, enhancing BERT's effectiveness for completeness checking of requirements.

\par\noindent\textbf{Keywords.} Requirements Completeness, Natural Language Processing (NLP), Machine Learning (ML), Large Language Models (LLMs), BERT.

\section{Introduction}\label{sec:introduction}

Natural language (NL) is widely used in industry to express systems and software requirements. Despite its prevalence, NL requirements are  prone to incompleteness. Improving the completeness of NL requirements is an important yet challenging problem in requirements engineering (RE)~\cite{Zowghi:03,Zowghi:03b}. 
The RE literature identifies two different notions of completeness~\cite{Zowghi:03b}: (1)~\emph{Internal} completeness pertains to requirements being closed in terms of the functions and qualities that can be deduced solely from the requirements. (2)~\emph{External} completeness focuses on ensuring that requirements encompass all the information suggested by external sources of knowledge. These sources can include individuals (such as stakeholders) or artifacts like higher-level requirements and existing system descriptions~\cite{arora2019empirical}. External completeness is a relative measure since the external sources may themselves be incomplete, or not all relevant external sources may be known~\cite{Zowghi:03b}. While external completeness cannot be defined in absolute terms, relevant external sources, when available, can be useful for detecting missing requirements-related information.

\begin{figure*}[!t]
\centering
\includegraphics[width=1\linewidth]{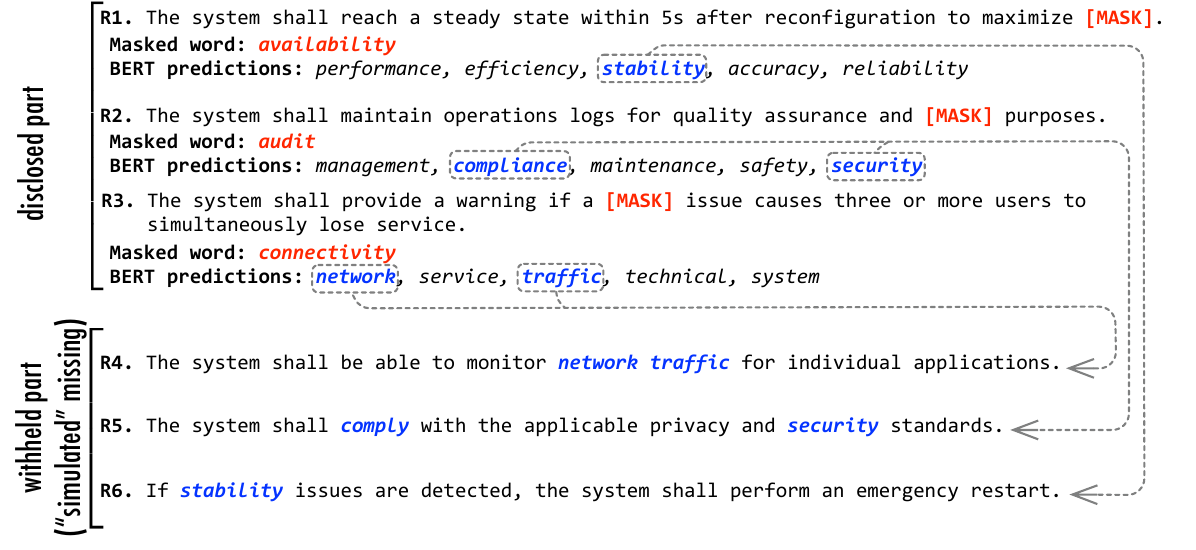}
\caption{Illustrative requirements specification split into a \emph{disclosed} and a \emph{withheld} part. The withheld part \emph{simulates} requirements omissions. Masking words in the disclosed part and having BERT make predictions for the masks reveals some terms that appear only in the withheld part.}
\label{fig:example}
\end{figure*}

\subsection{Motivation}
Natural Language Processing (NLP) is a powerful tool for computer-assisted verification of external completeness when requirements and external sources of knowledge are presented in text format. For example, Ferrari et al.~\cite{ferrari2014measuring} utilize NLP to assess the completeness of requirements by analyzing interview transcripts. In a similar vein, Dalpiaz et al.~\cite{dalpiaz2018pinpointing} integrate NLP with visualization techniques to identify disparities in stakeholders' perspectives. These disparities are in turn examined as potential indicators of incompleteness. 

Large language models (LLMs) provide a fresh opportunity to employ NLP for improving the external completeness of requirements. Through self-supervised learning, LLMs have been pre-trained on vast collections of textual data, e.g., millions of Wikipedia articles. This opens up the possibility of using LLMs as external knowledge sources for completeness checking. {\color{black}More precisely, we see two primary use cases, denoted as U1 and U2 below, for LLMs in relation to external completeness:

\begin{enumerate}
\item[U1] LLMs can be used for autonomously gathering, summarizing, and integrating new information from external sources, including their own pre-training data. 

\item[U2] LLMs can be tasked with suggesting alternative versions of existing material. While the results often do not convey the exact same meaning as the original material, they largely preserve context. An examination of the resulting variant material can reveal pertinent information that has been overlooked.
\end{enumerate}

In this article, we propose an instantiation of U2 (above) employing BERT (Bidirectional Encoder Representations from Transformers)~\cite{devlin2018bert} as our chosen LLM.}

BERT has been trained to predict masked tokens by finding words or phrases that most closely match the surrounding context. To illustrate how BERT can help realize U2, consider the example in Fig.~\ref{fig:example}. In this example, we have masked one word, denoted as \texttt{[MASK]}, in each of requirements $\textsf{R1}$, $\textsf{R2}$ and $\textsf{R3}$. We have then had BERT make five predictions for filling each masked slot. For instance, in $\textsf{R1}$, the masked word is \emph{availability}. The predictions made by BERT are: \emph{performance}, \emph{efficiency}, \emph{stability}, \emph{accuracy}, and \emph{reliability}. As seen from the figure, one of these predictions, namely \emph{stability}, is a word that appears in $\textsf{R6}$. Similarly, the predictions that BERT makes for the masked words in $\textsf{R2}$ and $\textsf{R3}$ (\emph{audit} and \emph{connectivity}) reveal new terminology that is present in $\textsf{R4}$ and $\textsf{R5}$ (\emph{network}, \emph{traffic}, \emph{comply} and \emph{security}).
In this example, \emph{if requirements $\textsf{R4}$--$\textsf{R6}$ were to be missing, BERT's predictions over $\textsf{R1}$--$\textsf{R3}$ would  provide useful cues about some of the missing concepts.}

\subsection{Contributions}\label{subsec:contributions}

{\color{black}The core utility of an LLM lies in its ability to predict and generate text that is both coherent and contextually relevant. This characteristic makes LLMs potentially useful tools for making recommendations on how to make requirements more complete. To systematically assess the usefulness of an LLM for this purpose, we need a strategy to evaluate the predictive accuracy of the LLM in identifying relevant content that is absent from requirements. To this end, we \emph{simulate} missing requirements information by randomly withholding a portion of a given requirements specification. We disclose the remainder of the specification to the LLM of choice, in our case BERT, to obtain predictions. Requirements in the disclosed portion are revealed one at a time to BERT for obtaining predictions of masked tokens. In our example of Fig.~\ref{fig:example}, the disclosed part would be requirements $\textsf{R1}$--$\textsf{R3}$, and the withheld part would be requirements $\textsf{R4}$--$\textsf{R6}$.}

{\color{black}When BERT is employed as a recommender in the manner described above, an important consideration is striking a trade-off between valuable and superfluous recommendations.
Our \emph{first contribution} is configuring BERT's number of predictions per mask to find a reasonable balance between the identification of simulated omissions vis-à-vis the generation of unhelpful predictions or noise.} We observe that achieving good coverage of requirements omissions through BERT predictions results in a significant amount of noise. Some of the noise can be easily filtered. For instance, in the example of Fig.~\ref{fig:example}, one can dismiss the predictions of \emph{service} and \emph{system} (made over \textsf{R3}); these words already appear in the disclosed portion, thereby providing no cues about missing terminology. Furthermore, one can dismiss words that carry little meaning, e.g., ``any'', ``other'' and ``each'', should such words appear among the predictions. After applying these obvious filters, the predictions still remain considerably noisy. Our \emph{second contribution} is a machine learning-based filter that post-processes predictions made by BERT, aiming to reduce the occurrence of noise in the predictions.

We based our solution development and evaluation on a set of 40 requirements specifications from the PURE dataset ~\cite{ferrari2017pure}. These specifications collectively comprise over 23,000 sentences. To support replication and enable future research, we have made our implementation and evaluation artifacts publicly available~\cite{luitel2023replication}.

{\color{black} Our evaluation results suggest that BERT's masked language model has the potential to assist in improving the completeness of requirements. Nevertheless, our current work does not attempt to build a user-facing tool for using BERT in requirements completeness checking, nor does it conduct user studies to measure practical benefits. While our findings are encouraging, they do not constitute conclusive evidence of usefulness but rather represent a necessary first step towards further investigations in the future.}

\textcolor{black}{This article extends a previous conference paper~\cite{luitel2023paper} published at the 29th Working Conference on Requirements Engineering: Foundation for Software Quality (REFSQ'23). Compared to the conference version, the present article offers enhancements to the background and related work, provides substantial new evaluation that includes a comparison with baselines, and features improvements to the implementation of our approach.}

\subsection{Organization}
Section~\ref{sec:background} provides background information. Section~\ref{sec:related} examines the related work, focusing on completeness checking of NL requirements and NLP techniques in RE. Section~\ref{sec:approach} presents our approach. Section~\ref{sec:eval} reports on our experimental design, analysis and results. \textcolor{black}{Section~\ref{sec:discussion} employs human feedback to validate a design choice in our evaluation regarding the matching of predictions with simulated omissions.} Section~\ref{sec:limitation} discusses the limitations of our approach and the validity of our findings.  Section~\ref{sec:conclusion} summarizes the article and suggests  directions for future research.

\section{Background}\label{sec:background}

Below, we review the background for our work, covering the NLP pipeline, large language models, word embeddings, machine learning and corpus extraction.

\subsection{NLP Pipeline} \label{subsec:NLP}
\begin{figure}[h]
    \centering    \includegraphics[width=0.45\linewidth]{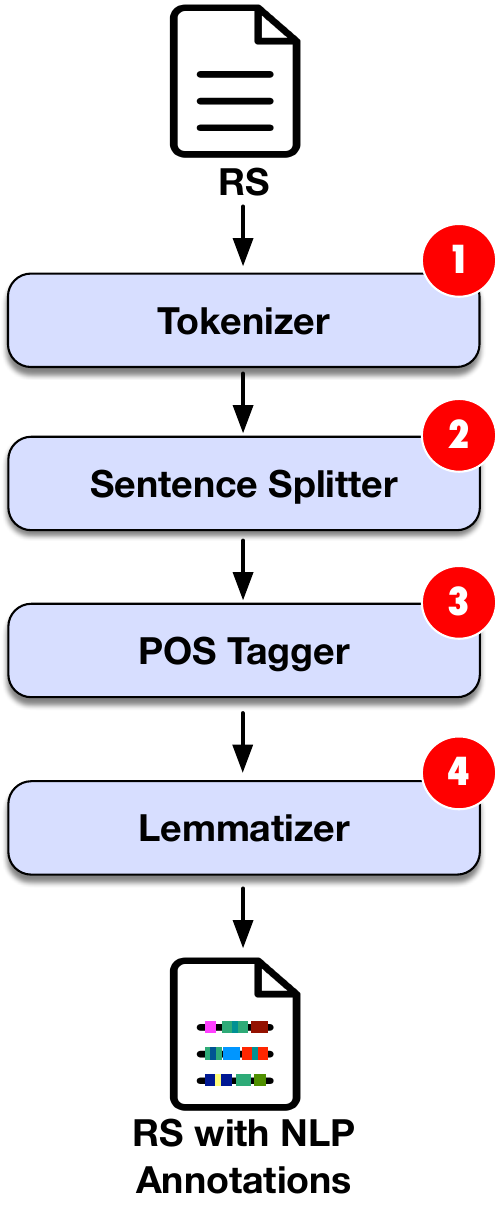}
    \caption{NLP Pipeline.}\label{fig:NLP_Pipeline}
    \end{figure}

    NLP is a branch of artificial intelligence (AI) that is concerned with automated analysis and representation of natural language, both text and speech~\cite{young2018recent}. NLP is usually performed using a pipeline of modules~\cite{hirschberg2015nlp}. The exact pipeline varies depending on the specific NLP task, and the tools and models used to perform it. The NLP pipeline we need in this article is presented in Fig.~\ref{fig:NLP_Pipeline}. We use the annotations produced by this pipeline for several purposes, including the identification and lemmatization of terms in requirements documents as well as processing predictions made by BERT in their surrounding context. 
     
    The modules in our NLP pipeline are as follows:
    
    \sectopic{Tokenizer.} The tokenizer breaks the text into individual words or tokens. Tokenization can be performed using various techniques such as whitespace-based, rule-based, or statistical methods~\cite{jurafsky2019speech}. For example, using whitespace, the sentence ``The model shall be implemented in Python.'' would be tokenized as: `the', `model', `shall', `be', `implemented', `in', `Python' and `.'.
    \sectopic{Sentence Splitter.} The sentence splitter breaks the text into individual sentences. 
    Sentence splitting typically uses punctuation marks, capitalization, and other cues to identify sentence boundaries~\cite{jurafsky2019speech}. It is important to note that the structure of what constitutes a sentence is predefined and may not necessarily correspond to grammatically correct sentences.
    
    \sectopic{Part-of-speech (POS) Tagger.} The POS tagger assigns to each word in each sentence a POS label, such as noun, verb, adjective or adverb. Continuing with our previous example, ``The model shall be implemented in Python.'', the POS tags assigned would be:
    `The':~DT (determiner),
    `model':~NN (singular noun),
    `shall':~MD (modal auxiliary verb),
    `be':~VB (base form verb),
    `implemented':~VBN (past participle verb),
    `in':~IN (preposition),
    `Python':~NNP (singular proper noun), and \linebreak
    `.':~PUNCT (punctuation). 
    
    \sectopic{Lemmatizer.} The lemmatizer reduces words to their base form, known as the lemma. This process helps normalize words with different inflections, allowing the pipeline to treat them as the same entity. Lemmatization improves text analysis by reducing vocabulary size and ensuring words with the same root meaning are treated equally. For example, the lemma for both `running' and `ran' is `run'.

\subsection{Large Language Models (LLMs)}\label{subsec:LM}
     A large language model (LLM) is a type of artificial-intelligence model designed specifically to understand and generate human language. LLMs are typically built using deep-learning techniques, particularly variants of neural networks such as transformers~\cite{devlin2018bert}. In this article, our LLM of choice is the Bidirectional Encoder Representations from Transformers (BERT)~\cite{devlin2018bert}. BERT is pre-trained using two self-supervised tasks: Masked Language Modelling (MLM) and Next Sentence Prediction (NSP). We use the MLM to identify closely related alternative words that may be relevant but are currently missing from an input RS. 
     
     MLM, or the Cloze task, is a procedure of randomly masking a percentage of tokens from a natural-language input and then attempting to predict the masked token~\cite{devlin2018bert}. When feeding an input text to BERT, there are three special tokens to take into consideration. `[CLS]' is a classification token appended to the start of every input to demarcate the beginning of the text. `[SEP]' is a separator token to mark the end of one sentence from the beginning of another. And, `[MASK]' is a masking token for the MLM task; `[MASK]' replaces an actual word in a sentence, prompting the prediction of contextualized tokens likely to match the masked word. Continuing with the example from Section~\ref{subsec:NLP}, the sentence ``The model shall be implemented in Python.'' would be modified into ``[CLS] The model shall be implemented in Python. [SEP]'' before tokenization. If we were to mask the word `Python', the sentence would be updated to ``[CLS] The model shall be implemented in [MASK]. [SEP]''. Examples of predictions made by BERT are provided over requirements \textsf{R1}-\textsf{R3} in Fig.~\ref{fig:example}. 
     
     BERT uses a bidirectional encoder for generating context-aware word representations. The encoder consists of a stack of Transformer blocks that use self-attention mechanisms~\cite{vaswani2017attention} to encode the input sequence in both directions. 
     BERT Base and BERT Large are two types of the BERT model. BERT Large, while generally more accurate, requires more computational resources. BERT Base has 12 encoder layers with a hidden size of 768, 12 self-attention heads, and $\approx$110 million trainable parameters. Further variations of BERT include cased and uncased models. For BERT uncased, the text has been lower-cased before tokenization, \linebreak whereas in BERT cased, the tokenized text is the same as the input text. Previous RE research suggests that the cased model is preferred over uncased for analyzing requirements~\cite{hey2020norbert,ezzini2022automated}. To both mitigate computation costs and follow best practices in RE, we employ the BERT-base-cased model for our experimentation. 
    
\subsection{Word Embeddings}\label{subsec:WE}
     In our work, predictions generated by MLM may have similar meanings in the context of a match against the ground truth, even if the terms themselves are not lexically identical. We thus need a semantic notion of similarity for matching predictions from the MLM against what is desired based on the ground truth. 
     For instance, we aim to establish matches between terms such as (i)~`key' and `unlock', and (ii)~`encryption' and `security', despite them not being lexically equivalent. For this, we use cosine similarity over \emph{word embeddings}. Cosine similarity quantifies the similarity of two words by calculating the distance between their vector representations~\cite{manning2008introduction}. 
     To obtain a vector representation, we must first transform each word into its own word embedding. Word embeddings are mathematical representations of words as dense numerical vectors capturing syntactic and semantic regularities~\cite{mikolov2013linguistic}. 
     
     To construct word embeddings, we use GloVe (Global Vectors for Word Representation), an unsupervised machine learning model that creates word embeddings using a co-occurrence matrix~\cite{pennington2014glove}. Different versions of GloVe have different dimensionality for embeddings. We use GloVe with a dimensionality of 50.
     Our decision to employ GloVe's pre-trained model for obtaining non-contextualized word embeddings~\cite{pennington2014glove} is motivated by striking a trade-off between accuracy and efficiency. BERT generates contextualized word embeddings; however, these embeddings are expensive to compute because they take context into consideration. BERT embeddings thus do not scale well when a large number of pairwise term comparisons is required, which happens to be the case in our evaluation.

\subsection{Machine Learning (ML)}\label{subsec:ML}
\textcolor{black}{%
Machine learning (ML) can be divided into three main types: unsupervised learning, supervised learning, and reinforcement learning. Unsupervised learning uses unlabelled data, which lacks predefined categories, to find patterns or relationships in the data without prior knowledge of the output labels. In supervised learning, a labelled dataset is used to learn the relationships between the input features and the output labels. Finally, reinforcement learning focuses on agents interacting with an environment and learning optimal actions through trial and error by receiving feedback in the form of rewards or penalties based on the consequences of their actions.}
In this article, we use supervised learning for distinguishing relevant from non-relevant predictions made by BERT. Our empirical evaluation (Section~\ref{sec:eval}) examines several widely used supervised ML algorithms, namely Neural Network (NN), Decision Tree (DT), Logistic Regression (LR), Random Forest (RF), and Support Vector Machine (SVM), in order to determine which one(s) are most accurate for our purpose.

    Classification techniques rely on the extraction of relevant features from the input data. The features can be nominal, numeric, or ordinal. Numeric features are continuous or discrete numerical values. Nominal features are discrete values that describe some categorical aspect of the data. Ordinal features have an inherent order through ranking to indicate a higher or lower value in relation to one another. Feature selection is the process of selecting a subset of relevant features from the original set of features to improve the accuracy and efficiency of a machine learning model~\cite{liu2012feature,cai2018feature}. The goal is to eliminate irrelevant or redundant features that do not contribute significantly to the model's performance, while retaining the most important features. We employ feature selection to rank the importance of features and ensure that only the most valuable ones are computed and retained. Our features for learning and our process for creating labelled data are discussed in Sections~\ref{sec:approach}~and~\ref{sec:eval}, respectively.

    Classification models have a tendency to predict the more prevalent class(es)~\cite{WittenFrankEtAl17}. Furthermore, classification algorithms typically give equal treatment to different misclassification types when minimizing misclassification. In many problems, however, the costs associated with different misclassification are not symmetric. In our context, non-relevant terms outnumber relevant ones. This increases the likelihood of terms being classified as non-relevant, thus increasing the risk of false negatives (i.e., useful terms being filtered out).
    We under-sample the majority class (i.e., non-relevant) to counter imbalance in the training set and thus reduce the risk of filtering useful information~\cite{berry2021hairy}. We assign a higher penalty to relevant terms being filtered than non-relevant terms being classified as relevant. In other words, we prioritize Recall over Precision which is often necessary in RE tasks. 
    A second strategy we employ in order to  prioritize Recall is cost sensitive learning (CSL). CSL can improve accuracy by assigning different misclassification costs to different classes or types of errors, with the aim of minimizing the prevalence of certain types of misclassified data~\cite{fernandez2018cost}. In our approach, filtering a false negative (i.e., ``relevant'' prediction) is more detrimental than filtering a false positive (i.e., ``non-relevant'' prediction). We thus use CSL to adjust the model's parameters and bias it towards reducing the prevalence of false negatives, even if it leads to a higher number of false positives, to ensure that useful predictions do not get filtered.

\subsection{Domain-corpus Extraction}\label{sec:DCE}
    Domain-specific corpora are useful resources for improving the accuracy of automation in RE~\cite{ezzini2021using}. The domain can be any specific subject or field, such as healthcare, telecommunications, or law. Domain-specific corpus extraction is the process of creating a corpus from existing texts or documents that are relevant to a particular domain. In this article, we use domain-specific corpora to enhance the accuracy of classifiers in filtering out non-relevant terms from BERT predictions.
    
    The first step in domain-corpus extraction is to define the domain's scope and determine relevant sources to the domain. If a domain-specific corpus already exists, there may be no need to generate a new one. 
    However, if a suitable corpus is not available, then one can be extracted manually or automatically. 
    Corpus creation can be based on domain documents from sources such as books, magazines, and online resources. 

    Due to its extensive coverage and diverse range of articles, Wikipedia is among the most common sources used for corpus extraction~\cite{ezzini2021using,cui2008corpus,ferrari2017detecting,ezzini2022wikidominer}. Manual corpus extraction involves searching for and selecting relevant texts or documents by hand. This process can be time-consuming and requires domain expertise to ensure that the selected texts are representative of the domain. Automated corpus extraction navigates a collection of textual documents according to pre-specified criteria to build a corpus. The criteria for extraction may be based on keywords, topic modelling, or named-entity recognition~\cite{ezzini2021using}. To avoid burdening users with manual tasks and minimize the cost of using our approach, we opt for automated corpus extraction. To do so, we use an existing tool, named \emph{WikiDoMiner}~\cite{ezzini2022wikidominer}. This tool has been specifically designed to generate domain-specific corpora by crawling Wikipedia.
    
    \begin{figure}[!t]
    \centering
        \includegraphics[width=\linewidth]{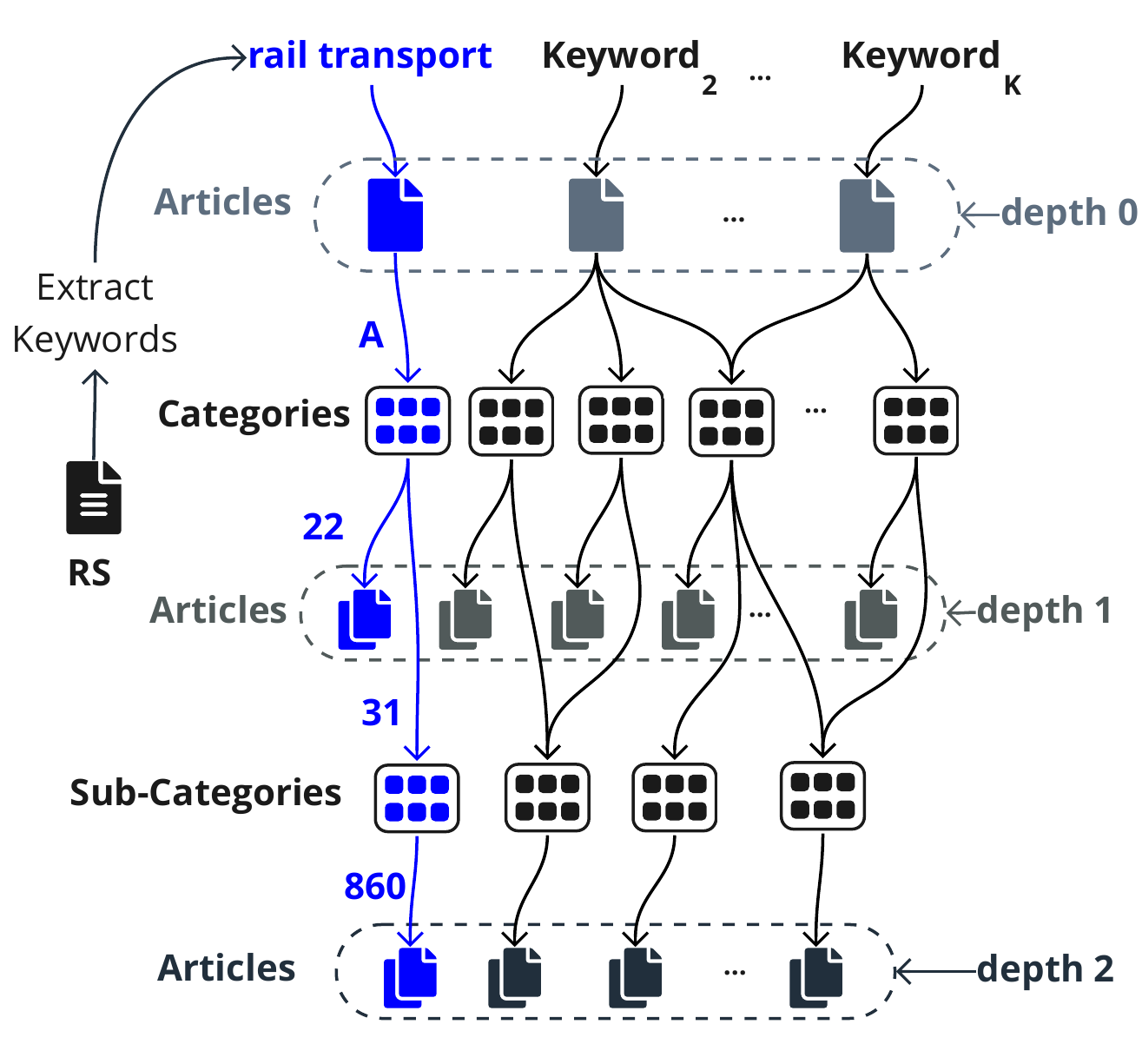}
    \caption{Traversing Wikipedia Categories~\cite{ezzini2022wikidominer}.}\label{fig:WikiDoMiner}
    \vspace*{-1.5em}
    \end{figure}
    
    WikiDoMiner allows control over corpus expansion through a depth parameter. A depth of zero creates a corpus of articles directly matching the key phrases in the input document, while increasing the depth results in larger corpora that encompass sub-categories of Wikipedia articles. To better understand how \linebreak WikiDoMiner queries Wikipedia, consider the scenario borrowed from Ezzini et al.~\cite{ezzini2022wikidominer} and illustrated in Fig.~\ref{fig:WikiDoMiner}. In Wikipedia, articles are categorized and can belong to multiple hierarchical categories. For instance, a search for the keyword ``rail transport'' in the Railway domain may yield an article titled ``Rail Transport.'' By analyzing the category structure of this article, we find that it falls under a category with the same name, referred to as Category A in Fig.~\ref{fig:WikiDoMiner}. Further exploration of a sub-category such as ``Rail Infrastructure'' reveals additional pages and sub-categories. As we explain in Section~\ref{sec:approach}, we use domain-specific corpora extracted by WikiDoMiner to calculate frequency-based \hbox{statistics for filtering.}

\section{Related Work}\label{sec:related}

In this section, we discuss and compare with pertinent areas of related work in the RE literature, focusing on  (1)~completeness checking of NL requirements, and (2)~NLP for requirements engineering. 

\subsection{Completeness Checking of NL Requirements}
There are various methods for determining the completeness of NL requirements. Espa\~{n}a et al.~\cite{espana2009completeness} measure completeness by comparing use cases against information systems through communication analysis. The authors evaluate the level of completeness achieved by reviewed models against a reference model. Gigante et al.~\cite{gigante2015verification} use ontological engineering to determine completeness of requirements. They model  requirements in the form of (subject, predicate, object) triplets. These triplets are then compared against an external source to verify completeness. 

Eckhardt et al.~\cite{eckhardt2016incompleteness} propose a framework consisting of a unified model for performance requirements and a content model to 
capture relevant content for addressing incompleteness. This process utilizes sentence patterns derived from the content model to evaluate completeness of requirements. Alrajeh et al.~\cite{alrajeh2012obstacle} create synthetic obstacles to verify completeness of requirements. The approach uses model checking and iteratively generates domain-specific obstacles, with the process ending after achieving a domain-complete set of obstacles. The above works cross validate requirements against an external model to measure completeness. Our approach seeks to address the same challenge by using a generative language model, BERT, and its vast pre-training data as a knowledge source for making contextualized predictions. Arora et al.~\cite{arora2019empirical} conduct a case study to detect external incompleteness of requirements using domain models. The authors simulate requirements omissions and demonstrate that UML class diagrams can display a near-linear sensitivity to detecting missing and under-specified requirements. Dalpiaz et al.~\cite{dalpiaz2018pinpointing} develop a technique based on NLP and visualization to explore commonalities and differences between multiple viewpoints and thereby help stakeholders pinpoint occurrences of ambiguity and incompleteness. Differences may occur when terms appear in a single viewpoint, i.e., the situation where a viewpoint refers to concepts that do not appear in other viewpoints. In the above works, the sources of knowledge used for completeness checking are existing development artifacts. Our approach does not require any user-provided artifacts. We leverage BERT's capacity to predict relevant terminology, independently of supplementary artifacts, to ensure domain independence. 

Bhatia et al.~\cite{bhatia2018semantic} address incompleteness in privacy policies by representing data actions as semantic frames. A semantic frame is constructed by identifying relevant questions for the data action, as semantic roles. Semantic roles represent the relationship of different clauses in statements to the main action. They identify the expected semantic roles for a given frame, and consequently determine incompleteness by identifying missing role values. Cejas et al.~\cite{cejas2021ai} use NLP and ML for completeness checking of privacy policies. Their approach identifies instances of pre-defined concepts such as ``controller'' and ``legal basis'' in a given policy. They create a conceptual model as a hierarchical representation of metadata types referring to different GDPR concepts based on hypothesis coding. It then verifies through rules whether all applicable concepts are covered. The above works deal with privacy policies only and have a predefined conceptual model for textual content. Our BERT-based approach is not restricted to a particular application domain and does not have a fixed conceptualization of the textual content under analysis. Instead, we utilize BERT's pre-training and attention mechanism to make contextualized recommendations for improving completeness. 

\subsection{NLP for Requirements Engineering}
Natural Language Processing for Requirements Engineering (NLP4RE) is a field that employs techniques from NLP to address challenges faced in the RE domain. Applications of NLP4RE include terminology extraction~\cite{hasso2022termextraction}, requirements similiarity and retrieval~\cite{abbas2021similarity}, user story analysis~\cite{lucassen2016userstory}, and legal requirements analysis~\cite{slemi2018legal}. 

The state of the art in NLP4RE has been extensively covered in a recent literature review by Zhao et al.~\cite{zhao2021nlp4re}. Two of the papers identified in this literature review utilize BERT-based language modelling, although neither work focuses on evaluating completeness of requirements. The first paper by Hey et al.~\cite{hey2020norbert} presents a new method for unsupervised representation learning, called NoRBERT (Non-functional
and functional Requirements classification using BERT). NoRBERT can be used for a wide range of NLP tasks, particularly when labeled data is limited or unavailable. The second paper by Sainani et al.~\cite{sainani2020contracts} uses BERT for automating the extraction and classification of requirements from software engineering contracts. Although both papers employ BERT, neither work applies BERT's MLM task in their approach.

In more recent literature, Shen and Breaux~\cite{shen2022domain} propose an NLP-based approach for extracting domain \linebreak knowledge from word embeddings and user-authored scenarios. Their approach involves gathering a corpus of scenarios authored by users in four distinct directory-service domains - apartments, hiking trails, restaurants, and health clinics. Then, the authors extract basic domain models from these scenarios by utilizing typed dependencies. The authors use seed question templates that include a domain-specific noun, seed verb and mask, and utilize MLM to predict substitute tokens for the mask. While this approach is not concerned with checking the completeness of requirements, it uses BERT's MLM for generating alternative entities by masking words in requirements statements. Our approach uses BERT's MLM in a similar manner. In contrast to the above work, we take steps to address the challenge arising from such use of BERT over requirements, namely the large number of non-relevant alternatives (false positives) generated. We propose a ML-based filter that uses a combination of NLP and statistics extracted from a domain-specific corpus to reduce the incidence of false positives.

\section{Approach}\label{sec:approach}
\begin{figure*}[!t]
\centering
    \includegraphics[width=1\linewidth]{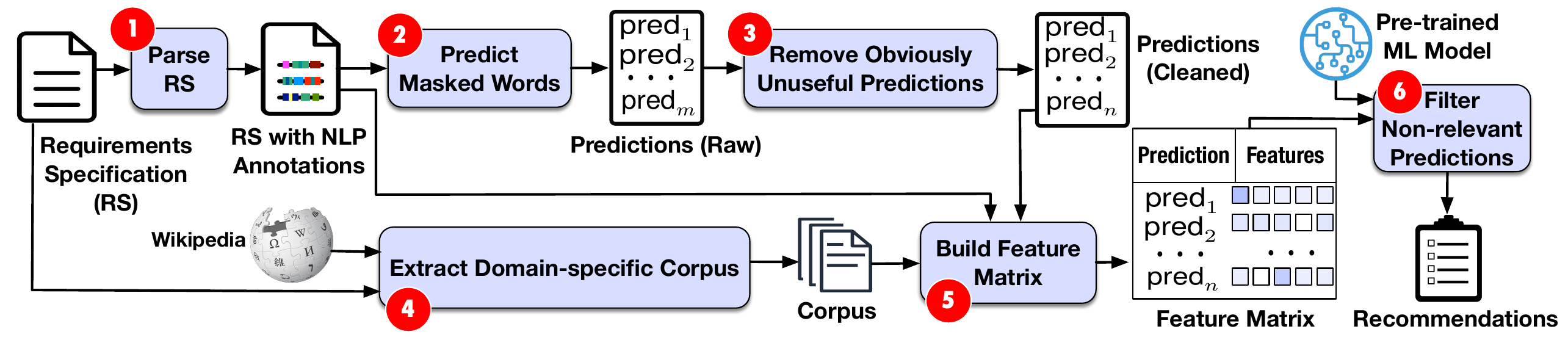}
\caption{Approach Overview.}\label{fig:approach}
\end{figure*}

Figure~\ref{fig:approach} provides an overview of our approach. The input to the approach is a (textual) requirement specification (RS). The approach has six steps. The first step is to parse the RS. The second step is to generate predictions for masked words using BERT. The third step is to remove the predictions that  provide little or no additional information.
The fourth step is to construct a domain-specific corpus for the given RS. Using this corpus and
the results from Step~1, the fifth step is to build a feature matrix for ML-based filtering of non-relevant terms from predictions by BERT. The sixth and last step is to feed the computed
feature matrix to a (pre-trained) classifier in an attempt to remove noise (non-relevant
words) from the predictions. The output of the approach is a list of
recommended terms that are likely relevant to the RS but are \hbox{currently absent from it.}

\subsection{Parsing RS using NLP}
To recommend relevant terms that may be missing from a given RS, we first apply to the RS the NLP pipeline presented in Section~\ref{sec:background}. Annotations resulting from this step include tokens, sentences, POS tags, and lemmas. Subsequent steps utilize the POS tags for masking and predicting terms.

\subsection{Obtaining Predictions from BERT}
The core meaning-bearing elements of requirements are \emph{nouns} and \emph{verbs}~\cite{Krzeszowski:11,arora2019empirical}. We iterate through each sentence in the annotated RS derived from Step~1. Based on the POS tags from Step~1, we  mask one noun or one verb at a time, creating a sentence with a single concealed word. Consider the sentence: ``The system shall generate reports on inventory levels, product movement, and sales history.'' The POS tags for the tokens in this sentence are:
`The'~(DT);
`system'~(NN);
`shall'~(MD);
`generate'~(VB);
`reports'~(NNS);
`on'~(IN);
`inventory'~(NN);
`levels'~(NNS);
`,'~(COMMA);
`product'~(NN);
`movement'~(NN);
`,'~(COMMA);
`and'~(CC);
`sales'~(NNS);
`history'~(NN);
`.'~(PERIOD).
%
For this sentence, examples of sentences with one concealed word are: ``The system shall [MASK] reports on inventory levels, product movement, and sales history'', and ``The system shall generate [MASK] on inventory levels, product movement, and sales history''. Each modified sentence is inputted to BERT, which generates a configurable number of predictions for the masked word. For example, in our illustration shown in Fig.\ref{fig:example}, we use BERT to generate five predictions per masked word. Based on our empirical evaluation  in Section~\ref{sec:eval}, we recommend using 15~predictions per masked word. BERT generates a probability score for each prediction, indicating its level of confidence. We retain the probability scores for use in Step~5 of our approach. 

\subsection{Removing Obviously Unuseful Predictions}
We attempt to improve the accuracy of our approach by removing predictions that are clearly not useful. The first category of predictions we discard are those that are already present in the RS. Consider the following sentence: ``The system shall provide a programmable interface to support system integration.'' If the word `integration' is masked and BERT happens to predict exactly the same term back, then we do not include the prediction in the prediction list, because it does not offer any new insights. We identify two further categories of predictions that are unlikely to contribute meaningfully to the final output: (1) predictions that fall within the top 250 most commonly used words in the English language, since these words are too generic, and (2) predictions that fall within vague words and stop words for requirements, as per the lists compiled by Berry et al.~\cite{Berry:03} and Arora et al.~\cite{arora2015tse}~\cite{arora2017tse}. The end result of this step is a more focused list of predictions that is cleared of obviously unhelpful terms.

\subsection{Generating Domain-specific Corpus for RS}
In Step~4 of our approach, we use WikiDoMiner (introduced in Section~\ref{sec:DCE}) to automatically extract a domain-specific corpus for an input RS~\cite{ezzini2022wikidominer}. 

For example, if an aerospace engineering RS is fed into WikiDoMiner, the resulting corpus may contain articles on aircrafts, aviation, and fluid dynamics. Recall that as the depth parameter of WikiDoMiner increases, the corpus grows larger, encompassing more sub-categories of Wikipedia articles at each level. In our work, we limit our search to direct article matches only (i.e., \hbox{\emph{depth~= 0}}). This decision was motivated by the following considerations: First, by limiting the depth to zero, we significantly reduce the volume of text being processed, enabling faster corpus generation. Second, focusing on direct article matches helps scope the expansion of terminology to the content that is immediately relevant to the domain of the input RS. During our exploratory investigation, we discovered that increasing the depth value 
results in diluting domain-specificity. 
Our decision to use a depth value of zero strikes a balance between corpus size and domain coverage, yielding better-suited corpora for our purposes. Step~5 uses the domain-specific corpus to compute features, which are subsequently employed for filtering non-relevant predictions.

\begin{table*}
\caption{Features for Learning Relevance and Non-relevance of Predictions Made by BERT.}
\label{tbl:featurestable}
\fontsize{10}{12}\selectfont
\begin{tabularx}{\textwidth}{p{1.5em} X}
  \toprule
  \textbf{ID} & 
  \textbf{Type (T), 
  Definition (D) and
  Intuition (I)}\\
  \midrule
  \hline
  F1 &
  \textbf{(T)} Nominal \textbf{(D)} POS tag of the masked word (noun or verb).  \textbf{(I)} This feature is helpful if nouns and verbs happen to influence relevance in different ways.\\

  F2 &
  \textbf{(T)} Nominal \textbf{(D)} POS tag of the prediction; this is obtained by replacing the masked word with the predicted word and running the NLP pipeline on the resulting sentence. \textbf{(I)} The intuition is similar to F1, except that predictions are not necessarily nouns or verbs and can, e.g., be adjectives or adverbs.\\
  
  F3 &
  \textbf{(T)} Nominal (Boolean) \textbf{(D)} True if F1 and F2 match; otherwise, False. \textbf{(I)} A mismatch between F1 and F2 could be an indication that the prediction is non-relevant. \\ 
 
  F4 &
  \textbf{(T)} Numeric \textbf{(D)} Length (in characters) of the masked word. \textbf{(I)} Words that are too short may give little information. As such, predictions resulting from masking short words could be non-relevant. \\ 

  F5 &
  \textbf{(T)} Numeric \textbf{(D)} Length (in characters) of the prediction. \textbf{(I)} Predictions that are too short could be non-relevant. \\ 

  F6 &
  \textbf{(T)} Numeric \textbf{(D)} $\min\text{(F4, F5)}/\max\text{(F4, F5)}$. \textbf{(I)} A small ratio (i.e., a large difference in length between the prediction and the masked word) could indicate non-relevance. \\
  
  F7 &
  \textbf{(T)} Numeric \textbf{(D)} The confidence score that BERT provides alongside the prediction. \textbf{(I)} A prediction with a high confidence score could have an increased likelihood of being relevant. \\ 

  F8 &
  \textbf{(T)} Numeric \textbf{(D)} Levenshtein distance between the prediction and the masked word. \textbf{(I)} A small Levenshtein distance between the prediction and the masked word could indicate relevance. \\ 
  
  F9 &
  \textbf{(T)} Numeric \textbf{(D)} Semantic similarity computed as cosine similarity over word embeddings. \textbf{(I)} A prediction that is close in meaning to the masked word could have a higher likelihood of being relevant. \\
  
  F10$^{*}$ &
  \textbf{(T)} Ordinal \textbf{(D)} A value between zero and nine, indicating how frequently the prediction (in lemmatized form) appears across \emph{all BERT-generated predictions} over a given RS. \textbf{(I)} A smaller value could indicate a higher likelihood of relevance.\\ 
  
  F11$^{*}$\,$^{\dagger}$ &
  \textbf{(T)} Ordinal \textbf{(D)} A value between zero and nine, indicating how frequently the prediction (in lemmatized form) appears in the \emph{domain-specific corpus}. \textbf{(I)} A smaller value could indicate a higher likelihood of relevance.\\ 
  
  F12$^{\dagger}$\,$^{\ddagger}$ &
  \textbf{(T)} Numeric \textbf{(D)} Average TF-IDF rank of the prediction across all articles in the domain-specific corpus. \textbf{(I)} A higher rank could indicate a higher likelihood of relevance. \\ 
  
  F13$^{\dagger}$\,$^{\ddagger}$ &
  \textbf{(T)} Numeric \textbf{(D)} Maximum TF-IDF rank of the prediction across all articles in the domain-specific corpus. \textbf{(I)} Same intuition as that for F12.\\ 
  \hline
  \bottomrule
\end{tabularx}

\vspace*{.5em}
$^{*}$Zero is most frequent (top ten percentile) and nine is least frequent (bottom ten percentile).
$^{\dagger}$Feature uses domain-specific corpus.
$^{\ddagger}$TF-IDF values are normalized by Euclidean norm.
\end{table*}

\subsection{Building Feature Matrix for Filtering}

For each prediction from Step~3, we compute a feature vector as input for a ML-based classifier that decides whether the prediction is ``relevant'' or ``non-relevant'' to the input RS. Our features are listed and explained in Table~\ref{tbl:featurestable}. 

The main principle behind our feature design has been to keep the features generic and in a normalized form. Being generic is important because we do not want the features to rely on any particular domain or terminology. Having the features in a normalized form is important for allowing labelled data from multiple documents to be combined for training, and for the resulting ML models to be applicable to unseen documents. The output of this step is a feature matrix where each row represents a prediction (from Step~3) and each column represents a feature as defined in Table~\ref{tbl:featurestable}.

Most of the features listed in Table~\ref{tbl:featurestable} can be easily understood based on the accompanying definitions. Below, we provide additional explanation for F10--F13 which are more involved. F10 and F11 are based on quantile bucketing. To categorize the frequency of predictions into discrete intervals or ``buckets'', we can divide the terms into equal-sized groups based on percentiles. Suppose that we have a bag of 1000 predictions generated by BERT, and that we wish to create 10 quantile buckets based on the frequency of these predictions. To do so, we first count the number of occurrences of each predicted word and sort them in descending order of frequencies. We then divide the words into 10 equally-sized groups based on their rank to create the quantile buckets. We assign the most frequently predicted words to bucket~0, and the least frequently predicted words to bucket~9. F12 and F13 are based on Term Frequency-Inverse Document Frequency (TF-IDF)~\cite{tf-idf} -- a common technique for measuring the importance of terms (words) in a particular domain. 

\subsection{Filtering Noise from Predictions}
The predictions from Step~3 are noisy (i.e., have many false positives). To reduce the noise, we subject the predictions to a pre-trained ML-based filter. The most accurate ML algorithm for this purpose is selected empirically (see RQ3 in Section~\ref{sec:eval}). The selected algorithm is trained on the development and training portion of our dataset ($P_1$ in Table~\ref{tbl:dataset}, as we discuss in Section~\ref{sec:eval}). Due to our features in Table~\ref{tbl:featurestable} being generic and normalized, the resulting ML model can be used as-is over unseen documents without re-training (see RQ4 in Section~\ref{sec:eval} for evaluation of effectiveness). The output of this step is the list of BERT predictions that are classified as ``relevant'' by our filter; duplicates are excluded from the final results.

\section{Evaluation}\label{sec:eval}
In this section, we empirically evaluate our approach. During the process, we also build the pre-trained ML model required by Step~6 of the approach (Fig~\ref{fig:approach}).

\subsection{Research Questions (RQs)}
Our evaluation answers the following RQs using part of the PURE dataset~\cite{ferrari2017pure}. In lieu of expert input about incompleteness for the documents in this dataset, we apply the withholding strategy discussed in Section~\ref{subsec:contributions} to simulate incompleteness.

\sectopic{RQ1. How accurately can BERT predict relevant but missing terminology for an input RS?} The number of predictions generated by BERT per mask is a configurable parameter. RQ1 identifies the optimal value offering the best balance for producing useful recommendations. This investigation will focus on the optimal number of predictions per mask between the range of 5 and 20. Choosing an optimal number of predictions is essential for ensuring that our approach maintains the potential to provide benefits. If the number of predictions is too low, the approach may miss out on relevant recommendations. Conversely, if the number of predictions is too high, the approach may generate too much noise, thus making it difficult for users to identify the most relevant recommendations.
{\color{black}%
\sectopic{RQ2. How does our approach compare to baselines?} RQ2 examines whether the contextualized recommendations made by BERT are advantageous over recommendations that one can obtain through simpler means. To this end, we conduct a comparative analysis of the quality of predictions generated by our approach against three baseline methods.}

\sectopic{RQ3. Which ML classification algorithm most accurately filters unuseful predictions made by BERT?} Useful recommendations from BERT come alongside a considerable amount of noise. RQ3 investigates different ML algorithms for filtering this noise. The RQ further explores the impact of data balancing and cost-sensitive learning to mitigate over-filtering.

\sectopic{RQ4. How accurate are the recommendations generated by our approach over unseen documents?} In RQ4, we combine the best BERT configuration from RQ1 with the top-performing filter models from RQ3, and measure the accuracy of this combination over unseen data.

\subsection{Implementation and Availability}\label{sec:implementation}
\begin{figure}[!t]
\centering
    \includegraphics[width=0.85\linewidth]{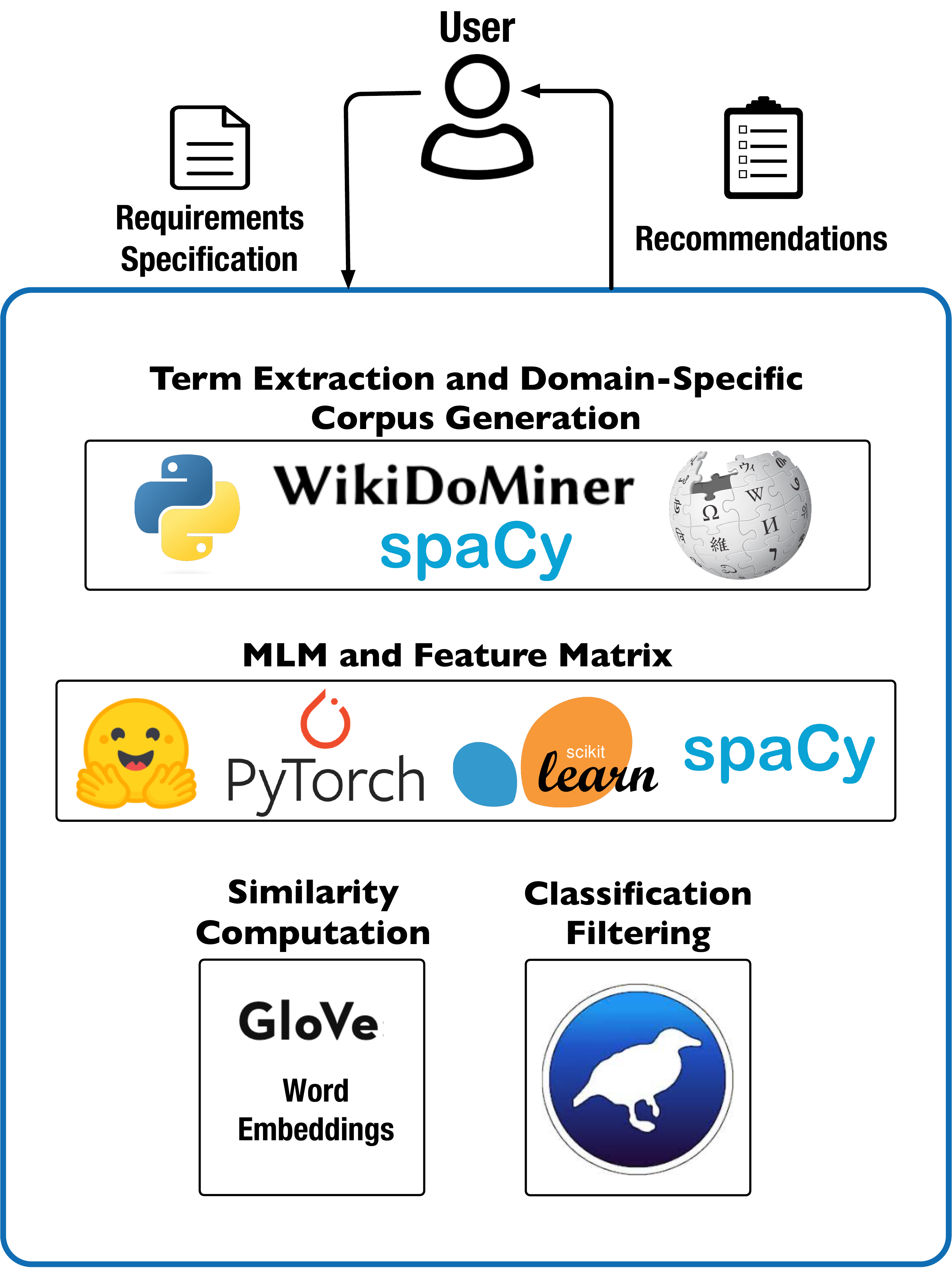}
\caption{Tool Overview.}\label{fig:Tool_Overview}
\end{figure}

Figure~\ref{fig:Tool_Overview} presents an overview of the implementation of our approach. The implementation is mostly in Python. Our NLP pipeline is implemented using SpaCy version 3.2.2. To extract word embeddings, we use GloVe~\cite{pennington2014glove}. For obtaining masked language model predictions from BERT, we use the Transformers library version 4.16.2 by Hugging Face (\url{https://huggingface.co/}), operated in PyTorch version 1.10.2+cu113. Our ML-based filters are implemented using WEKA 3-8-5~\cite{frank2016weka} -- a lightweight tool for data mining and knowledge discovery.  
We use standard implementations of Levenshtein distance and cosine similarity (over word embeddings) to implement features F8 and F9 of Table~\ref{tbl:featurestable}, respectively. The TF-IDF-based features in Table~\ref{tbl:featurestable}, namely F12 and F13, are calculated using the TfidfVectorizer from scikit-learn version 1.0.2. The corpus required for the TF-IDF-based features is automatically constructed using the WikiDoMiner tool~\cite{ezzini2022wikidominer}. In addition, WordNet~\cite{wordnet} is used for discovering synonyms in one of our baselines (Baseline 3 as we discuss under EXPII in Section~\ref{subsec:proc}). All our implementation and evaluation artifacts are publicly available~\cite{luitel2023replication}.

\subsection{Dataset}\label{subsec:dataset}
Our evaluation is based on the PURE (Public Requirements) dataset~\cite{ferrari2017pure}. PURE is a collection of 79 publicly available  requirements documents, containing approximately 34,000 sentences. In our evaluation, we use 40 out of the 79 documents in PURE. Many of the documents in PURE require manual cleanup (e.g., removal of table of contents, headers, section markers, etc.). We found 40 to be a good compromise between the effort needed to spend on cleanup and having a dataset that is large enough for statistical significance testing, mitigating the effects of random variation, and training ML-based filters. The selected documents, listed in Table~\ref{tbl:dataset}, cover 15 domains.
We partition the documents into two (disjoint) subsets $P_1$ and $P_2$. $P_1$ is used for answering RQ1, RQ2, and RQ3; and, $P_2$ is used for answering RQ4.
Our procedure for assigning documents to $P_1$ or $P_2$ was as follows: We first randomly selected one document per domain and put it into $P_2$; this is to maximize domain representation in RQ4. From the rest, we randomly selected 20 documents for inclusion in $P_1$, while attempting to have $P_1$ represent half of the data in terms of token count. Any remaining document after this process was assigned to $P_2$, thus giving us 20 documents in $P_2$ as well. Table~\ref{tbl:dataset} provides domain information and summary statistics for \hbox{documents in $P_1$ and $P_2$ after cleanup.}

\begin{table}[!t]
\centering
    \caption{Our Dataset (Subset of PURE~\cite{ferrari2017pure}). $P_1$ is for development and training and $P_2$ for testing.}
   \includegraphics[width=\linewidth]{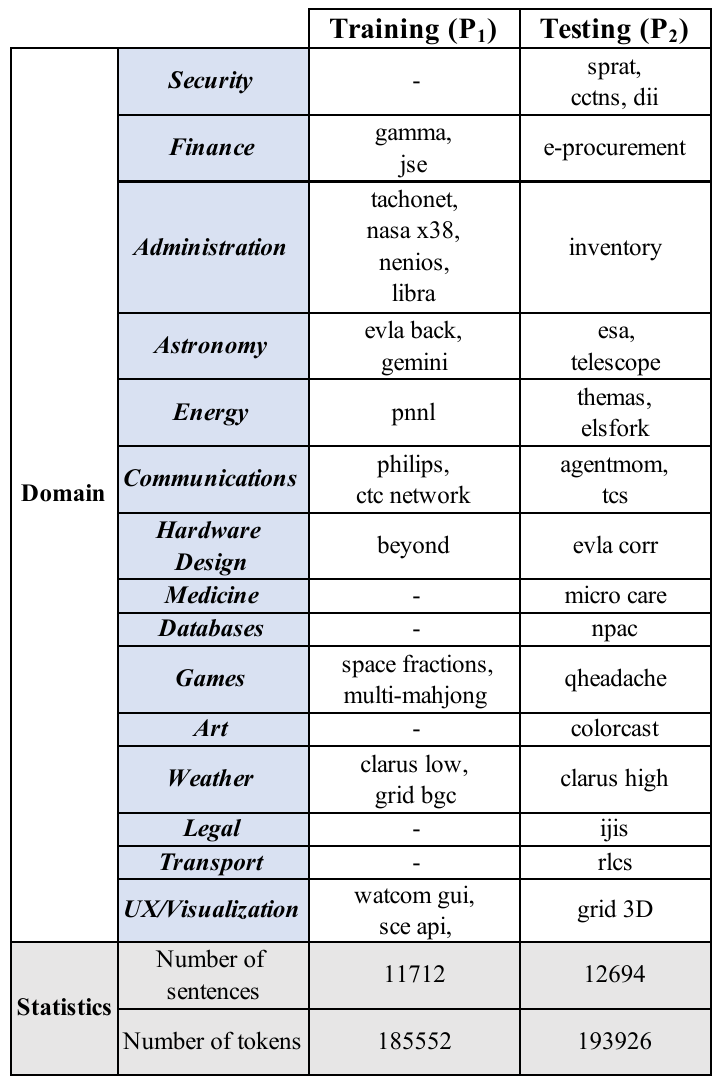}
    \label{tbl:dataset}
    \vspace*{-2.5em}
\end{table}

\subsection{Analysis Procedure}\label{subsec:proc}

\sectopic{EXPI.} This experiment answers RQ1. For every document $p\in P_1$, we randomly partition the set of sentences in $p$ into two subsets of (almost) equal sizes. In line with our arguments in Section~\ref{sec:introduction}, we \emph{disclose} one of these subsets to BERT and \emph{withhold} the other. We apply Steps~1,~2~and~3 of our approach (Fig.~\ref{fig:approach}) to the disclosed portion, \emph{as if this portion were the entire input document}. We run Step~2 of our approach with four different numbers of predictions per mask: 5, 10, 15, and 20. For every document $p\in P_1$, we compute two metrics, \emph{Accuracy} and \emph{Coverage}, defined in Section~\ref{subsec:metrics}. As the number of predictions per mask increases from 5 to 20, the predictions made by BERT reveal more terms that are relevant to the withheld portion. Nevertheless, as we will see in Section~\ref{subsec:results}, the benefits diminish beyond 15 predictions per mask.

To ensure that the trends we see as we increase the number of predictions per mask are not due to random variation, we pick different shuffles of each document $p$ across different numbers of predictions per mask. For example, the disclosed and withheld portions for a given document $p$ when experimenting with 5 predictions per mask are different random subsets than when experimenting with 10 predictions per mask.

\sectopic{EXPII.} This experiment answers RQ2. To ensure that our approach is worthwhile, we need to compare it to baselines. Since our approach uses an LLM as its external source of knowledge, we cannot compare it to existing external completeness checking approaches which require additional artifacts such as domain models or interview transcripts; such artifacts are not available for our requirements dataset. To demonstrate that our approach is worthwhile, we define three common-sense baselines that are not based on LLMs and, at the same time, do not require any purpose-built artifacts. The first baseline, Baseline~1, is the list of \textcolor{black}{250-to-2000} most common words in the English language. This baseline enables us to assess whether BERT leads to contextualized predictions that have greater specificity than generic recommendations. \textcolor{black}{The second baseline, Baseline~2, uses TF-IDF obtained over the domain-specific corpus generated in Step~4 of our approach. Specifically, the baseline recommends all terms with a TF-IDF score exceeding a predetermined threshold, which we have set at 0.01.} This baseline captures the most relevant terminology in a domain. By comparing against Baseline~2, we examine whether BERT's context-specific predictions have an advantage over the domain-specific but non-contextualized terms derived from a corpus. 
The third baseline, Baseline~3, collects  WordNet synonyms of the words found in the  disclosed portion of a given RS. Comparing with Baseline~3 allows us to assess whether the predictions generated by BERT extend beyond synonyms that can be obtained through simpler methods than an LLM. Appendix~\ref{sec:pseudo} outlines our three baselines in pseudo-code form. To facilitate a direct comparison with EXPI results, we apply the three baselines to $P_1$ and evaluate their performance using the \emph{Accuracy} and \emph{Coverage} metrics (defined in Section~\ref{subsec:metrics}).

\sectopic{EXPIII.} This experiment answers RQ3 and further constructs the training set for the ML classifier in Step~6 of our approach (Fig.~\ref{fig:approach}). We recall the disclosed and withheld portions as defined in EXPI. For every document $p\in P_1$, we label predictions as ``relevant'' or ``non-relevant'' using the following procedure: Any prediction matching some term in the withheld portion is labelled ``relevant''. The criterion for deciding whether two terms match is a cosine similarity of $\geq85$\% over GloVe word embeddings (introduced in Section~\ref{sec:background}). All other predictions are labelled ``non-relevant''. The threshold of 85\% allows only terms with the same lemma or with very high semantic similarity to be matched. 
In Section~\ref{sec:discussion}, we empirically justify the chosen threshold, ensuring its conservative nature to minimize superfluous matches.

For each prediction, a set of features is calculated as detailed in Step~5 of our approach. It is paramount to note that Step~4, which is a prerequisite to Step~5, \emph{exclusively} uses the content of the disclosed portion without any knowledge of the withheld portion. The above process produces labelled data for each $p\in P_1$. We aggregate all the labelled data into a single \emph{training set}. This is possible because our features (listed in Table~\ref{tbl:featurestable}) are generic and normalized. This process ensures that the ML-based filter is trained on a diverse range of data, allowing it to effectively filter out non-relevant predictions and yield more accurate results.

Equipped with a training set, we compare five ML algorithms: Feed Forward Neural Network (NN), Decision Tree (DT), Logistic Regression (LR), Random Forest (RF)  and Support Vector Machine (SVM). All algorithms are tuned with optimal hyperparameters that maximize classification accuracy over the training set. For tuning, we apply multisearch hyperparameter optimization using random search~\cite{bergstra2012random}. The basis for tuning and comparing algorithms is ten-fold cross validation. We experiment with under-sampling the ``non-relevant'' class with and without cost-sensitive learning (CSL); the motivation is reducing false negatives (i.e., relevant terms incorrectly classified as ``non-relevant''). For CSL, we assign double the cost (penalty) to false negatives compared to false positives (i.e., noise). We further assess the importance of our features using information gain (IG)~\cite{WittenFrankEtAl17}. In our context, IG measures how efficient a given feature is in discriminating ``non-relevant'' from ``relevant'' predictions. A higher IG value implies a higher discriminative power.

\vspace*{.5em}\sectopic{EXPIV.} This experiment answers RQ4 by applying our end-to-end approach to unseen requirements documents, i.e., $P_2$. To conduct EXPIV, we need a pre-trained classifier for Step~6 of our approach (Fig.~\ref{fig:approach}). This classifier needs to be  independent of $P_2$. We build this classifier using the training set derived from $P_1$, as discussed in EXPIII. {\color{black}EXPIV follows the same strategy as in EXPI, which is to randomly withhold a portion of each document $p$ (now in $P_2$ rather than in $P_1$) and attempting to predict the novel terms of the withheld portion. In contrast to EXPI, in EXPIV, predictions made by BERT are post-processed by a filter to reduce noise. Furthermore, whereas EXPI split each document into two roughly equal portions, i.e., a 50-50 split, EXPIV considers two different split ratios: a 50-50 split similar to EXPI, as well as a 90-10 split, where 90\% of the document is disclosed and only 10\% is withheld. The 50-50 split enables us to examine the usefulness of our approach where there is major incompleteness (i.e., half of the content is missing). The 90-10 split represents the situation where there is minor incompleteness. 

We repeat EXPIV \emph{five times} for each $p \in P_2$. This mitigates random variation resulting from the random selection of the disclosed and withheld portions, thus yielding more realistic ranges for performance. In EXPIV, we study three levels of filtering for the two split ratios considered. Noting that there are 20 documents in $P_2$, the results reported for EXPIV are based on $20\,\text{(documents)}$ $*$ $5\,\text{(repetitions)}$ $*$ $3\,\text{(filtering levels)}$ $*$ $2\,\text{(split ratios)} = 600$ runs of our approach.}

\subsection{Metrics}\label{subsec:metrics}
We define separate metrics for measuring (1) the quality of term predictions and (2)~the performance of filtering. The first set of metrics is used in RQ1, RQ2 and RQ4; and, the second set is used in RQ3.
To define our metrics, we need to introduce some notation. Let \hbox{$\mathsf{Lem}:\textsf{bag}\rightarrow \textsf{bag}$} be a function that takes a bag of words and returns another bag of words by \emph{lemmatizing} every element in the input bag. Let $\mathsf{U}: \textsf{bag}\rightarrow \textsf{set}$ be a function that removes duplicates from a bag and returns a set. Let $C$ denote the set of common words and stopwords as explained under Step~3 in Section~\ref{sec:approach}. Given a document $p$ treated as a bag of words, the terminological content of $p$'s disclosed portion, denoted $h_1$, is given by set  $X=\mathsf{U}(\mathsf{Lem}(h_1))$. In a similar vein, the terminological content of $p$'s withheld portion, denoted $h_2$, is given by \hbox{set $Y=\mathsf{U}(\mathsf{Lem}(h_2))$}. What we would like to achieve through BERT is to predict as much of the \emph{novel} terminology in the withheld portion as possible. This novel terminology can be defined as set \hbox{$N = (Y - X) - C$}. Let bag $V$ be the output of Step~3  (Fig.~\ref{fig:approach}) when the approach is applied \emph{exclusively} to the disclosed portion of a given document (i.e., $h_1$). Note that $V$ is already free of any terminology that appears in the disclosed portion, as well as of \hbox{all common words and stopwords.}

\sectopic{Quality of term predictions.}
Let set $D$ denote the (duplicate-free) lemmatized predictions that have the potential to hint at novel terminology in the withheld portion of a given document. Formally, let \linebreak $D= \textsf{U}(\textsf{Lem}(V))$. We define two metrics, \emph{Accuracy} and \emph{Coverage} to measure the quality of $D$.
\emph{Accuracy} is the ratio of terms in $D$ matching some term in $N$, to the total number of terms in $D$. \textcolor{black}{That is, $\mathit{Accuracy} = |\{t \in D \text{ s.t. } t\ \text{matches some}\ t' \in N\}|/|D|$.} A term $t$ matches another term $t'$ if the word embeddings have a cosine similarity of $\geq 85\%$ (already discussed under EXPIII in Section~\ref{subsec:proc}).
The second metric, \emph{Coverage}, is defined as the ratio of terms in $N$ matching some term in $D$, to the total number of terms in $N$. \textcolor{black}{That is, $\mathit{Coverage} = |\{t \in N \text{ s.t. } t\ \text{matches some}\ t' \in D\}|/|N|$.} The intuition for Accuracy and Coverage is the same as that for the standard Precision and Recall metrics, respectively. Nevertheless, since our matching is inexact and based on a similarity threshold, it is possible for more than one term in $D$ to match an individual term in $N$. Coverage, as we define it, excludes multiple matches, providing a measure of how much of the novel terminology in the withheld portion \hbox{is hinted at by BERT.}

\sectopic{Quality of filtering.}
As explained earlier, our filter is a binary classifier to distinguish relevance and non-relevance for the outputs from BERT. To measure filtering performance, we use the standard metrics of \emph{Classification Accuracy}, \emph{Precision} and \emph{Recall}. True positive (TP), false positive (FP), true negative (TN) and false negative (FN) are defined as follows: A TP is a classification of ``relevant'' for a term that has a match in set $N$ (defined earlier). A FP is a classification of ``relevant'' for a term that does \emph{not} have a match in $N$. A TN is a classification of ``non-relevant'' for a term that does not have a match in $N$. A FN is a classification of ``non-relevant'' for a term that does have a match in $N$. \emph{Classification Accuracy} is calculated as $(TP+TN)/(TP+TN+FP+FN)$. \emph {Precision} is calculated as $TP/(TP+FP)$ and \emph{Recall} as $TP/(TP+FN)$. \textcolor{black}{We note that the Classification Accuracy metric defined for filtering is distinct from the Accuracy metric defined for term predictions by BERT, and it is consistently referred to as such to prevent ambiguity.}

\subsection{Results}\label{subsec:results}

\begin{figure}[!t]
  \centering
  \begin{subfigure}[b]{0.45\textwidth}
    \includegraphics[width=1\linewidth]{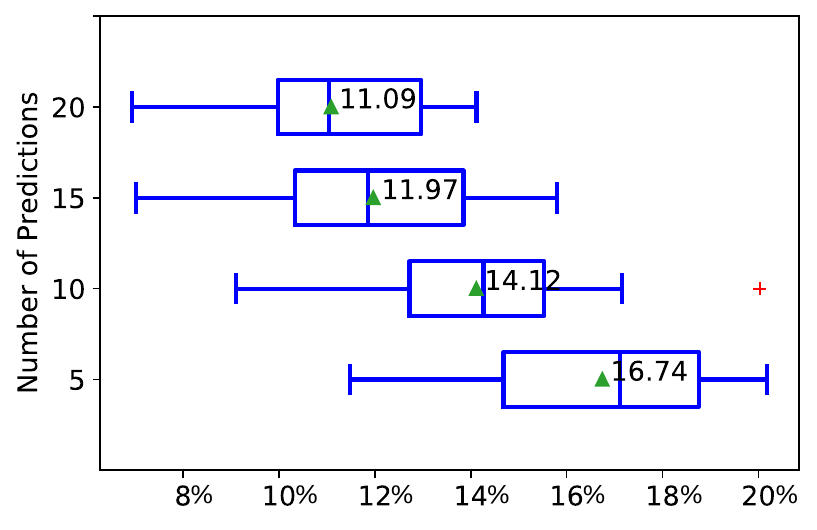}
    \caption{Accuracy.}\label{fig:RQ1_Accuracy}
      \vspace*{1em}
  \end{subfigure}
  \begin{subfigure}[b]{0.45\textwidth}
\includegraphics[width=1.05\linewidth]{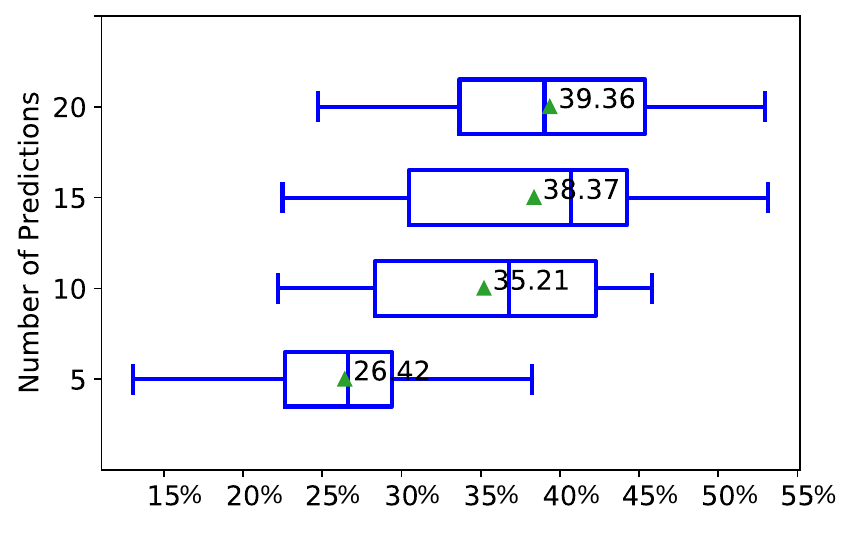}
  \caption{Coverage.}\label{fig:RQ1_Coverage}
  \end{subfigure}
  \caption{(a) Accuracy  and (b) Coverage for Different Numbers of Predictions per Mask}
  \label{fig:test1}
  \vspace*{-1.5em}
\end{figure}

\sectopic{RQ1.} Figures~\ref{fig:RQ1_Accuracy} and~\ref{fig:RQ1_Coverage} provide boxplots for Accuracy and Coverage with the number of predictions by BERT ranging from 5 to 20 in increments of 5. Each boxplot is based on 20 datapoints; each datapoint represents one document in $P_1$. 

We perform statistical significance tests on the obtained metrics using Wilcoxon's rank sum test~\cite{capon1991elementary} and Vargha-Delaney's $\hat{A}_{12}$~\cite{vargha2000critique}. 
{\color{black}
Wilcoxon's rank-sum test is a non-parametric statistical test used to assess whether there is a significant difference between the distributions of two independent samples by comparing the ranks of the observations. Vargha-Delaney's $\hat{A}_{12}$ is a non-parametric effect size measure that quantifies the magnitude of the difference between two groups. This metric measures the likelihood that an observation from one group is greater than an observation from another group. In our case, each group represents performance readings from a given approach for producing term recommendations. The $\hat{A}{12}$ value ranges from 0 to 1, where 0.5 indicates no difference between the two groups. Values less than 0.5 suggest the second group tends to have higher values, while values greater than 0.5 suggest the first group tends to have higher values. For instance, if $\hat{A}{12}$ equals 0.8 when comparing the performance of two approaches, $A$ and $B$, it implies that there is an 80\% probability that a randomly chosen performance result from $A$ will have a higher value than a randomly chosen performance result from $B$, suggesting a notable advantage of $A$ over $B$. The $\hat{A}_{12}$ is typically categorized as \emph{negligible}, \emph{small}, \emph{medium}, or \emph{large} based on the computed numeric value. Negligible signifies a very small difference, small implies a modest difference, medium indicates a moderate difference, and large denotes a substantial difference. We apply widely used thresholds, as suggested by Hess and Kromrey~\cite{hess2004robust}, \hbox{to derive these categories.}
}

Table~\ref{tbl:RQ1_Significance_Testing} shows the results of the statistical tests. Each row in the table compares Accuracy and Coverage across two levels of predictions per mask. For example, the \emph{5 vs. 10} row compares the metrics for when BERT generates 5 predictions per mask versus when it generates 10.

\begin{table}[!t]
\caption{Statistical  Tests for the Results of Fig.~\ref{fig:test1}.}
\label{tbl:RQ1_Significance_Testing}
\centering\includegraphics[width=0.85\linewidth]{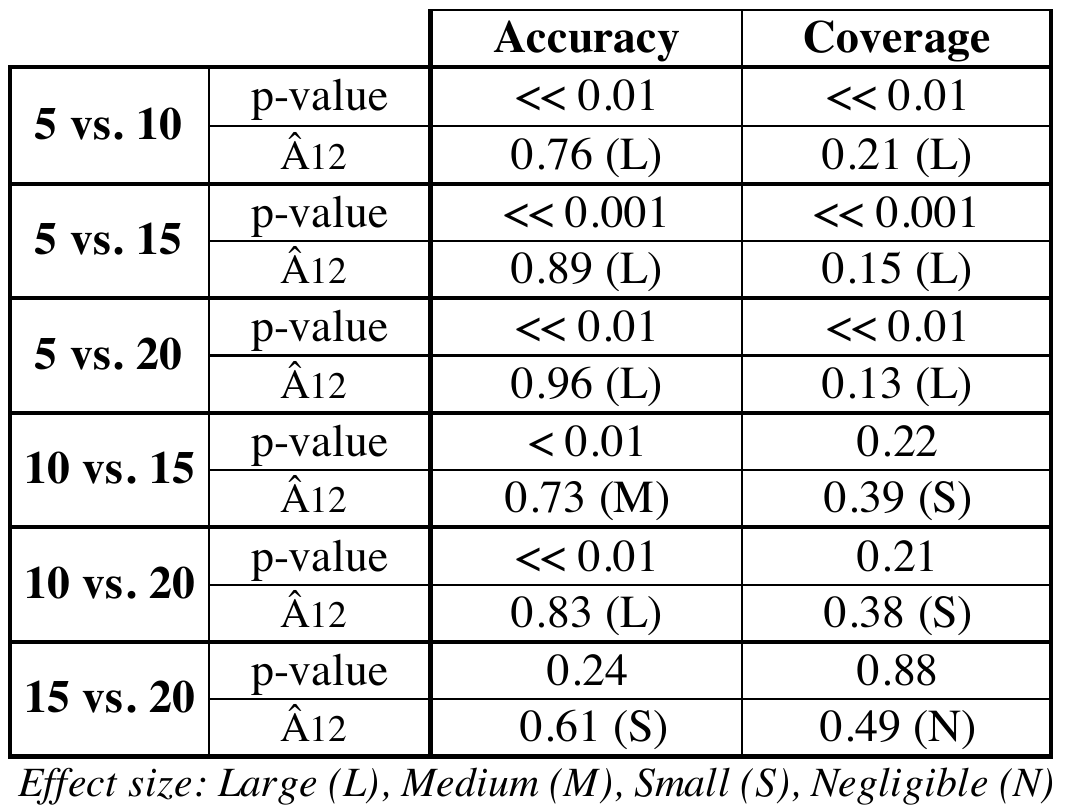}
\vspace*{-1em}
\end{table}

For Accuracy, Fig.~\ref{fig:RQ1_Accuracy} shows a downward trend as the number of predictions per mask increases. Based on Table~\ref{tbl:RQ1_Significance_Testing}, the decline in Accuracy is statistically significant with each increase in the number of predictions, the exception being the increase from 15 to 20, where the decline is not statistically significant. For Coverage, Fig.~\ref{fig:RQ1_Coverage} shows an upward but saturating trend. Five predictions per mask is too few: all other levels are significantly better. Twenty is too many, notably because of the lack of a significant difference for Coverage in the \hbox{\emph{10 vs. 20}} row of Table~\ref{tbl:RQ1_Significance_Testing}. The choice is thus between 10 and 15. We select 15 as this yields an average increase of 3.2\% in Coverage compared to 10 predictions per mask. This increase is not statistically significant. Nevertheless, the price to pay is an average decrease of $(14.12 - 11.97 =)\ 2.15\%$ in Accuracy. Given the importance of Coverage, we deem 15 to be a better compromise than 10.

\vspace*{1em}
\begin{samepage}
\begin{mdframed}[style=MyFrame]
\emph{The answer to {\bf RQ1} is: When requirements omissions are simulated by withholding, having BERT make 15 predictions per mask is the best trade-off for detecting missing terminology. BERT predicts terms that, on average, hint at $\approx$4 out of 10 omissions (Coverage $\approx$38\%). On average, $\approx$1 in 8 predictions is relevant (Accuracy $\approx$12\%).}
\end{mdframed}
\end{samepage}
\vspace*{.5em}
\vspace*{1em}


\begin{figure}[!t]
  \centering
  \begin{subfigure}[b]{0.475\textwidth}
    \includegraphics[width=\textwidth]{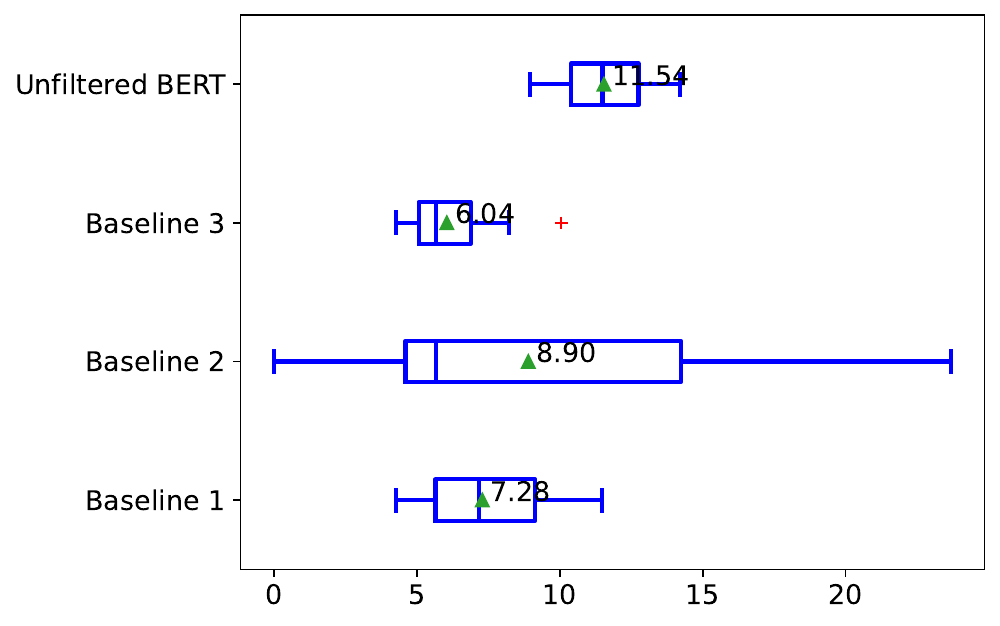}
    \caption{Accuracy. }\label{fig:RQ2_Accuracy}
    \vspace*{1.5em}
  \end{subfigure}
  \begin{subfigure}[b]{0.475\textwidth}
    \includegraphics[width=\textwidth]{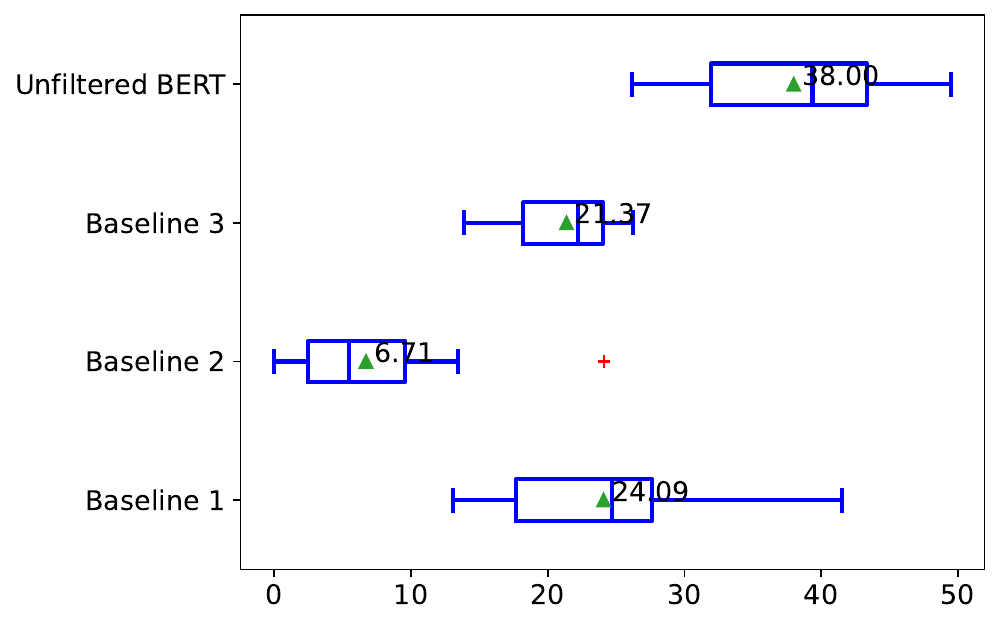}
    \caption{Coverage.}\label{fig:RQ2_Coverage}
  \end{subfigure}
  \caption{(a) Accuracy and (b) Coverage of Baselines Compared to BERT}
  \label{fig:test2}
\end{figure}

\begin{table}[!t]
\caption{Statistical Tests for the Results of Fig. \ref{fig:test2}.}
\label{tbl:RQ2_Significance_Testing}
\centering
\includegraphics[width=.9\linewidth]{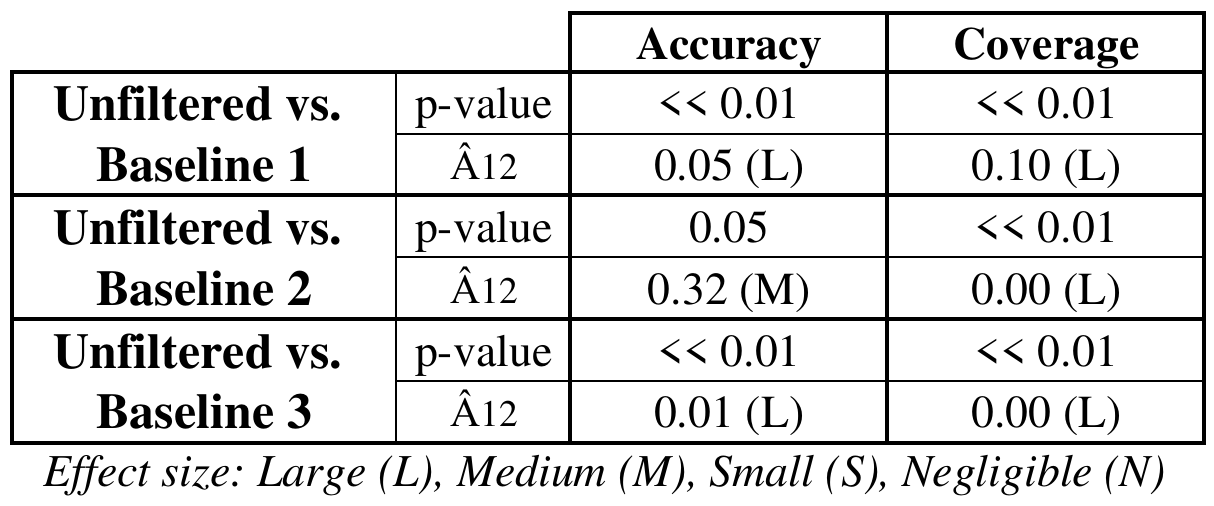}
\vspace*{-1em}
\end{table}

\sectopic{RQ2.} Figures~\ref{fig:RQ2_Accuracy} and \ref{fig:RQ2_Coverage} respectively compare the Accuracy and Coverage of the three baselines described in Section~\ref{subsec:proc} (EXPII) against the best BERT configuration identified in RQ1 (i.e., with 15 predictions made per mask). 
Statistical significance testing results are provided in Table~\ref{tbl:RQ2_Significance_Testing}. \textcolor{black}{
BERT outperforms all three baselines in terms of both Accuracy and Coverage. Two remarks need to be made regarding the baseline results: First, the Coverage of Baselines 1 and 2 can be increased by respectively increasing the number of most common words and adjusting the TF-IDF cutoff threshold (see EXPII in Section~\ref{subsec:proc}).
However, doing so entails a trade-off, as it leads to a decline in Accuracy for these baselines. Second, while the inclusion of Baselines~1 and 2 is important for benchmarking BERT's performance, a fundamental distinction exists between these two baselines and both our BERT-based solution and Baseline 3. BERT and Baseline 3 can trace their recommendations to the elements in the input requirements document. In contrast, Baseline~1 produces the same results regardless of the input requirements document. And, as for Baseline~2, there is no direct and easily interpretable connection between the recommendations and the requirements from which they originated. Although this difference is not reflected in the Accuracy and Coverage results, it is still significant for developing a practical solution. In real-world scenarios, engineers are unlikely to be interested in reviewing a list of recommendations without an explanation regarding the basis for these recommendations.}

\vspace*{1em}

\begin{samepage}
\begin{mdframed}[style=MyFrame]
\emph{The answer to {\bf RQ2} is: 
\textcolor{black}{
None of the three baselines discussed in Section~\ref{subsec:proc} present a better alternative to BERT due to a significant Coverage deficit coupled with lower Accuracy.}}
\end{mdframed}
\end{samepage}
\vspace*{1em}

\begin{table*}[t]
    \caption{ML Algorithm Selection (RQ3). All algorithms have tuned hyperparameters.}
    \label{fig:RQ3_Algorithms_Table}
   \includegraphics[width=\linewidth]{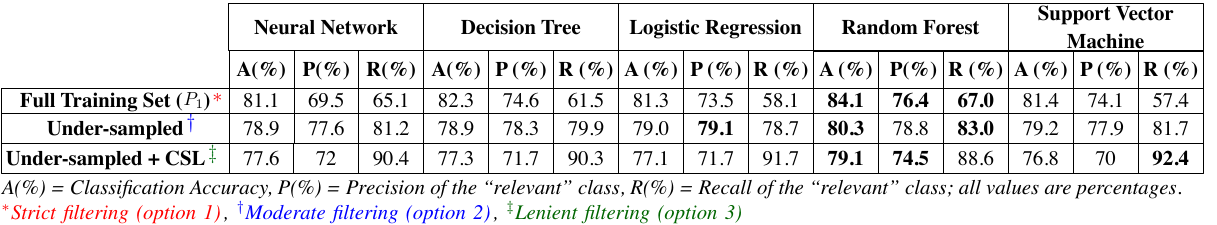}
\vspace*{.5em}
\end{table*}

\begin{figure}[htp]
  \centering
  \includegraphics[width=0.83\linewidth]{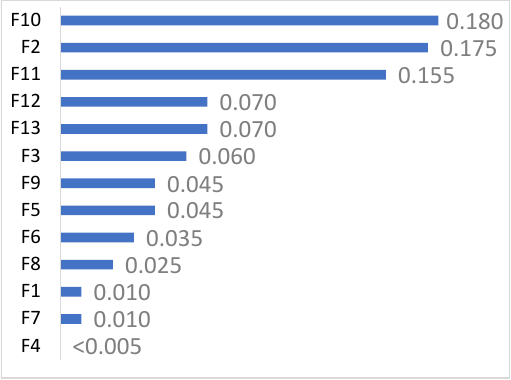}
  \caption{Feature Importance (Avg).}
\label{fig:Information Gain}
\end{figure}

\sectopic{RQ3.} Table~\ref{fig:RQ3_Algorithms_Table} shows the results for ML-algorithm selection using the full ($P_1$) training set (61,996 datapoints), the under-sampled training set (36,842 datapoints), and the under-sampled training set alongside CSL. Classification Accuracy, Precision and Recall are calculated using ten-fold cross validation. In the table, we highlight the best result for each metric in \textbf{bold}. When one uses the full training set (\emph{option~1}) or the under-sampled training set without CSL (\emph{option~2}), Random Forest (RF) turns out to be the best alternative. When the under-sampled training set is combined with CSL (\emph{option~3}), RF still has the best Accuracy and Precision. However, Support Vector Machine (SVM) presents a moderate advantage in terms of Recall. Since option~3 is meant at further improving the filter's Recall, we pick SVM as the best alternative for this particular option. Figure~\ref{fig:Information Gain} lists the features of Table~\ref{tbl:featurestable} in descending order of information gain (IG), averaged across options~1, 2 and 3. We observe that our corpus-based features (F11--F13) are among the most influential features, thus justifying the use of a domain-specific corpus extractor in our approach.

Compared to option~1, options~2~and~3 get progressively more lax by filtering \emph{less}. We answer RQ4 using RF for options 1 and 2 and SVM for option 3. For better intuition, we refer to option~1 as \emph{strict}, option~2 as \emph{moderate} and option~3 as \emph{lenient}.

\vspace*{.5em}
\begin{samepage}
\begin{mdframed}[style=MyFrame]
\emph{The answer to {\bf RQ3} is: RF and SVM yield the most accurate filter for unuseful predictions. RF is a better alternative for more aggressive filtering, whereas SVM is a better alternative for more lax filtering (thus better preserving Recall).}
\end{mdframed}
\end{samepage}
\vspace*{.5em}


\sectopic{RQ4.} {\color{black}In RQ1-RQ3, we applied a 50-50 split strategy to tune our approach over the training portion of our data.
In RQ4, we examine how effective our approach is over our test set, i.e., $P_2$ in Table~\ref{tbl:dataset}. For RQ4, as noted in EXPIV (Section~\ref{subsec:proc}), we consider both a 50-50 split strategy as well as a 90-10 strategy, respectively capturing major and minor incompleteness.

Table~\ref{fig:RQ4} shows boxplot results for different levels of filtering (unfiltered, lenient, moderate and strict) organized by 50-50 and 90-10 split strategies. We recall from EXPIV that five different random shuffles are performed for each $p\in P_2$. Each plot in Table~\ref{fig:RQ4} is therefore based on $5*20=100$ datapoints.

Without filters and over our test set, we observe an average Coverage of 40.04\% for the 50-50 split strategy and average Coverage of 39.25\% for the 90-10 strategy. The different split strategies represent remarkably different situations. Comparatively, the 90-10 strategy has 90/50=1.8 times more textual data to generate predictions on but the incompleteness is merely 10/50, i.e., one fifth, of the situation in the 50-50 split. That is, our approach has been capable of predicting approximately 40\% of the missing terminology, irrespectively of the level of incompleteness. There is nonetheless a notable difference in Accuracy between the 50-50 and 90-10 splits, with Accuracy for the former split strategy standing at 12.11\% and for the latter -- at 2.25\%. The lower Accuracy with a 90-10 split is explained by a combination of a higher number of predictions and a limited number of targets, stemming from the limited extent of incompleteness.}

\begin{table*}
\caption{Accuracy and Coverage over Test Set (RQ4).}\label{fig:RQ4}
\centering
\includegraphics[width=\linewidth]{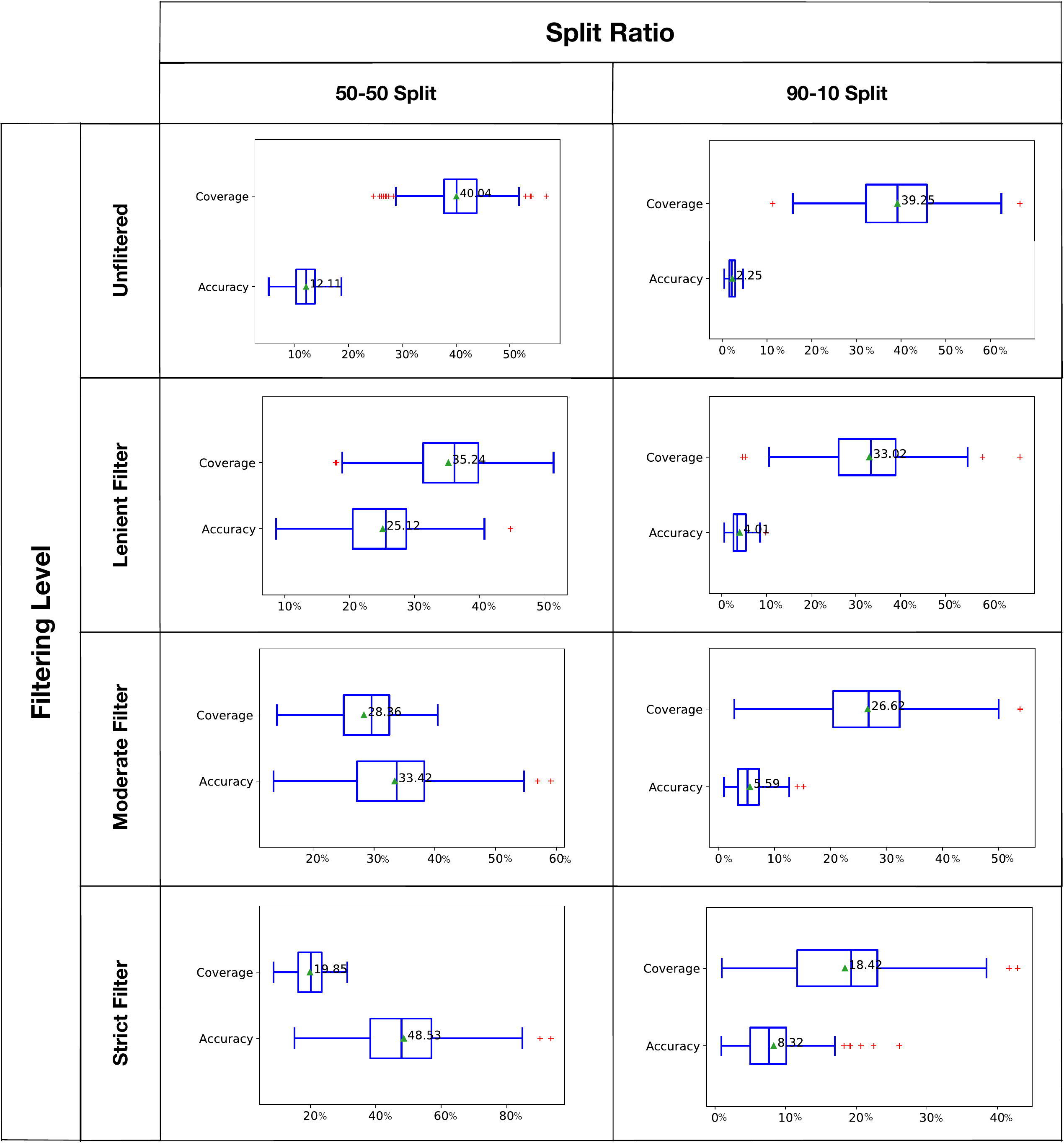}
\end{table*}

{\color{black}
Ultimately, which filtering option the user selects depends on how the user wishes to balance the overhead of reviewing non-relevant recommendations against potentially finding a larger number of relevant terms missing from requirements.

For the 50-50 split strategy, the lenient filter increases Accuracy by an average of $\approx$13\% while decreasing Coverage by $\approx$5\%. The moderate filter increases Accuracy by an average of $\approx$21\% while decreasing Coverage by $\approx$12\%. And, the strict filter increases Accuracy by an average of $\approx$36\% while decreasing Coverage by $\approx$20\%. As shown in Table~\ref{tbl:RQ4_Significance_Testing}, in the case of the 50-50 split, the strict and moderate filters increase Accuracy and decrease Coverage in a statistically significantly way and with large effect sizes. The lenient filter, on the other hand, increases Accuracy significantly and with a large effect size, while not negatively impacting Coverage in a statistically significant manner.

\begin{table*}[!t]
  \caption{Statistical Tests for the Results of Table~\ref{fig:RQ4}.}
  \label{tbl:RQ4_Significance_Testing}
  \centering
  \begin{subtable}[b]{0.475\textwidth}
    \caption{50-50 Split.}
    \includegraphics[width=\textwidth]{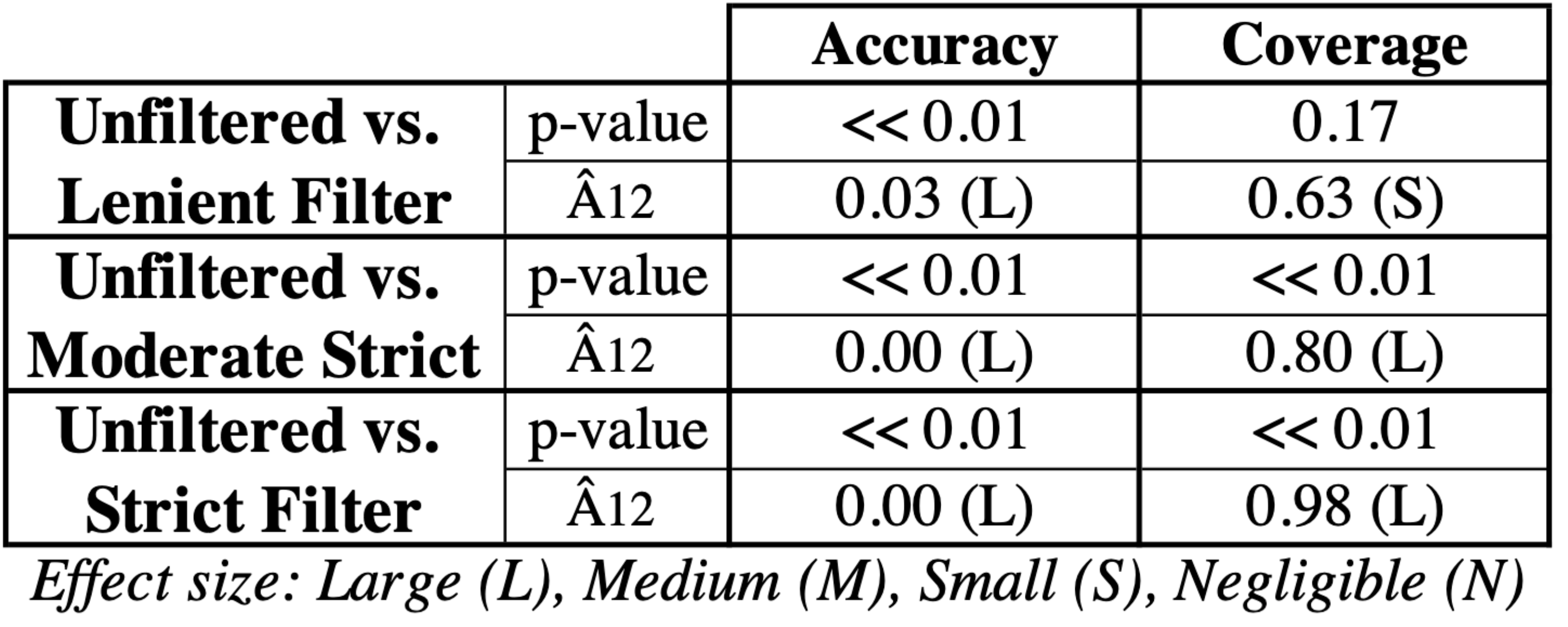}
  \end{subtable}
  \begin{subtable}[b]{0.475\textwidth}
    \caption{90-10 Split.}
    \includegraphics[width=\textwidth]{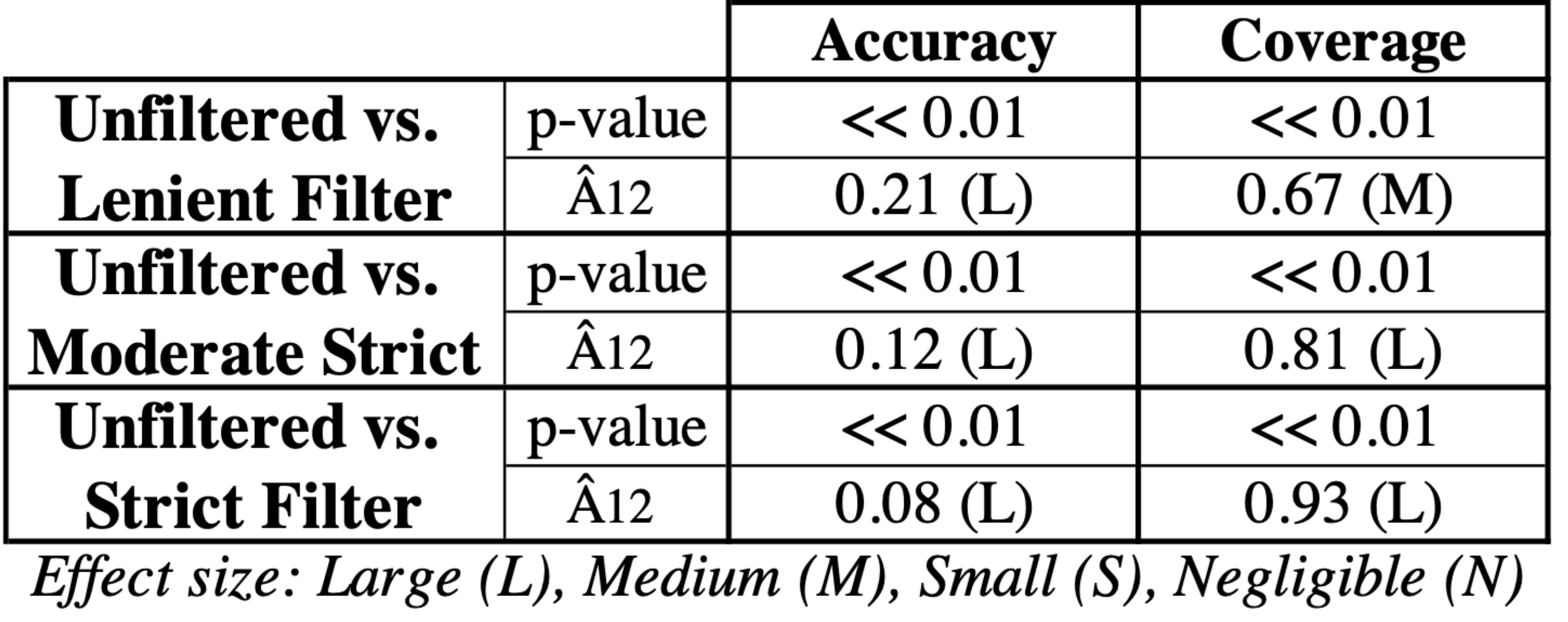}
  \end{subtable}
\end{table*}

As for the 90-10 split, the lenient, moderate, and strict filters increase Accuracy by an average of $\approx$1.76\%, $\approx$3.34\%, and $\approx$6.07\%, respectively. While increments in accuracy may appear small, it is important to note that false positives are significantly more prevalent than true positives when the anticipated amount of incompleteness is small. As such, the accuracy \emph{ratio} for different filtering levels is a better indication of how useful filtering is. For example, consider the Accuracy ratio between the lenient filter and the unfiltered data, which is calculated as $4.41/2.25 = 1.96$. This ratio signifies that, despite a relatively modest improvement in Accuracy (i.e., $4.41 - 2.25 =$ 1.76\%), the lenient filter effectively reduces the volume of terms requiring manual inspection by about \emph{half}, thus making filtering worthwhile.

In the 90-10 split strategy, the increase in Accuracy and the decrease in Coverage are statistically significant when filters are applied. As shown in Table~\ref{tbl:RQ4_Significance_Testing}, all effect sizes are large with one exception: the decline  in Coverage brought about by the lenient filter vs. the unfiltered has a medium effect size. This suggests that the lenient filter has a less severe impact on Accuracy compared to both moderate and strict filtering.

If one takes the preservation of Coverage as the main deciding factor, the lenient filter would be the best trade-off. When the anticipated amount of incompleteness is large (represented by the 50-50 split strategy), the lenient filter considerably improves Accuracy without a major impact on Coverage. And, when the anticipated amount of incompleteness is small (represented by the 90-10 split strategy), the lenient filter removes a substantial number of false negatives (around half) while decreasing Coverage with a moderate  effect size.

}

\begin{samepage}
\begin{mdframed}[style=MyFrame]
\it\color{black}
We answer \textbf{RQ4} in two distinct scenarios: (S1) when a significant degree of incompleteness is expected, represented by withholding 50\% of the content in a requirements document, and (S2) when a minor level of incompleteness is anticipated, represented by withholding 90\% of the content.

\mbox{}\color{black}For S1, when applying a strict filter, $\approx$48\% of recommendations from our approach are relevant, compared to $\approx$25\% with a lenient filter. Under lenient filtering, our recommendations provide cues for  $\approx$35\% of the (simulated) missing terminology. This number decreases to 20\% with a strict filter. As for S2 -- a more challenging scenario than S1 due to a larger number of recommendations and substantially fewer target terms for incompleteness -- unfiltered recommendations still offer clues for $\approx$39\% of the missing terminology. Filtering remains crucial for S2 as it eliminates a significant number of non-relevant terms; however, the prevalence of non-relevant terms remains high despite filtering. Further, due to the much smaller number of target terms compared to S1, the negative impact of filtering on relevant recommendations is more pronounced. This results in uncovering $\approx$33\%, $\approx$27\%, and $\approx$18\% of missing terminology for lenient, moderate, and strict filtering, respectively.
\end{mdframed}
\end{samepage}

\section{Discussion}\label{sec:discussion}
As explained in Section~\ref{subsec:proc} (EXPIII) and our metric discussion in Section~\ref{subsec:metrics}, we employ an 85\% cosine similarity threshold on word embeddings to assess whether predictions by BERT are good matches for the novel terms in the withheld half of a requirements document. This method presents an advantage over exact lexical matches by considering the semantic similarity between predictions and novel terms, thereby resulting in more thorough and nuanced matches. At the same time, it is important to exercise caution when determining the threshold, as setting it too low can result in erroneously considering dissimilar terms as valid matches, thus potentially negatively impacting the reliability of our evaluation. We selected the 85\% threshold based on an exploratory analysis in our earlier conference paper~\cite{luitel2023paper}, where we surmised that the chosen threshold would be conservative enough to rule out the majority of dissimilar matches. In this article, we seek to offer empirical evidence for our chosen threshold and quantify its influence on the interpretation of our findings.

We structure our discussion in two stages. First, we determine what proportion of the matches admitted by this threshold are exact lexical matches. Exact matches do not require human validation to assess their quality. In the second stage, we collect human feedback on the quality of a selected subset of non-exact matches. The goal here is to systematically estimate the percentage of these matches that provide useful insights to humans.

\begin{figure}[!t]
\centering
    \includegraphics[width=1\linewidth]{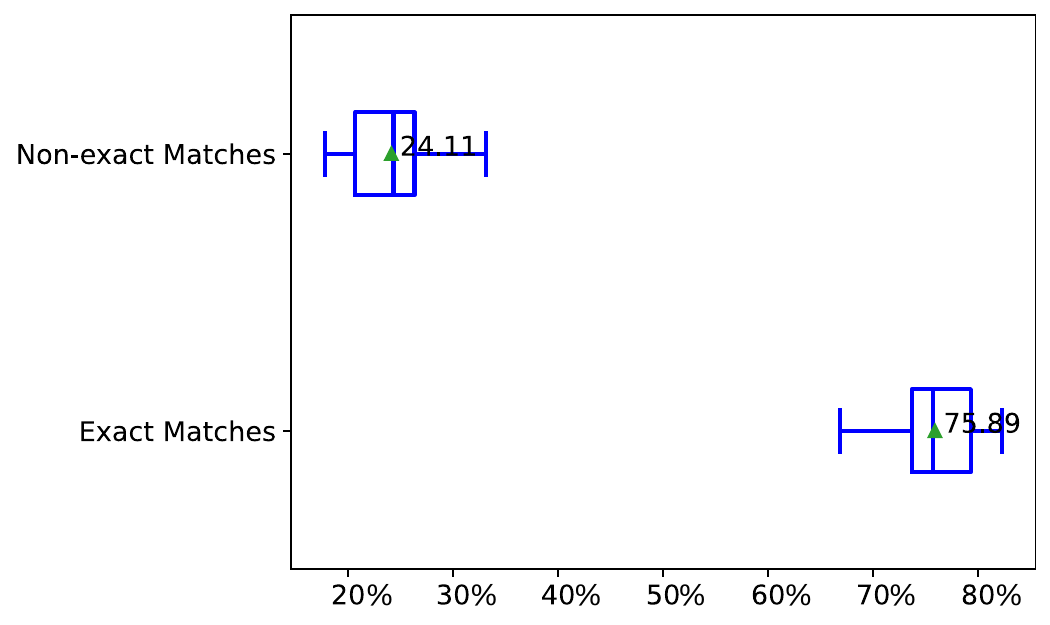}
\caption{Proportions of Exact vs. Non-exact Matches over Test Set.}\label{fig:matches}
\vspace*{-1.75em}
\end{figure}

Figure~\ref{fig:matches} shows the proportions of exact versus non-exact matches. Each data point represents a single run of our approach over each document in the test set ($P_2$), without any filters applied. The results indicate that the majority of matches are exact, with less than a quarter of them being non-exact. To assess the quality of non-exact matches, we engaged two non-authors to obtain unbiased human feedback. Both individuals are PhD students in Computer Science with excellent proficiency in English, over four years of industry experience, and prior exposure to requirements and requirements engineering. We selected 40 non-exact-match examples at random from our experimentation with the test set. These examples can be found in our supplementary material~\cite{luitel2023replication}. The two individuals were interviewed separately, allowing us to assess interrater agreement. The question that an individual had to answer for each example was as follows: \emph{For this given sentence, is term X (predicted by BERT) a useful match for term Y (which appears in the sentence)?} 

The notion of ``useful'' is subjective. To determine which predictions were considered useful, we relied on individual discretion while encouraging discussion to support the decisions made. Before each interview, we reviewed the criteria for usefulness. In particular, we clarified that a useful match could come in two ways: (1)~a synonym that conveyed the same meaning as the original word, or (2)~a term that would prompt reflection and deeper thought, such as a more specific term like ``car'' being provided for a more general original term like ``vehicle'' or vice versa.

The first individual found 75\% (30/40) of the non-exact predictions to be useful, while the second individual found 87.5\% (35/40) of the predictions to be useful. Out of the 40 examples shown to the individuals, there were only four cases (i.e., 10\%) where both individuals considered the non-exact prediction to be unuseful. Cohen's Kappa ($\kappa$) was used to measure the agreement level among the individuals. The result was $\kappa=0.44$, indicating moderate agreement. This implies that determining usefulness involves a non-negligible degree of subjectivity, which aligns with previous research in RE highlighting the subjective nature of concepts like relevance and usefulness~\cite{arora:2019active}. At the same time, we observe that depending on how one aggregates the human feedback, between 75\% and 90\% of the non-exact matches are useful. Noting that, on average, approximately 76\% of the matches are exact (Fig.~\ref{fig:matches}), the unuseful cases are a proportion of the remaining average 24\%. We thus estimate the error arising from non-exact matching to range from 2.4\% ($24\% \times 10\%$) to 6\% ($24\% \times 25\%$). Despite this margin of error, we consider the trade-off of employing non-exact matching worthwhile, given that, based on our results, useful matches outnumber unuseful ones by a ratio of at least three to one.

\section{Limitations and Validity Considerations}\label{sec:limitation}
In this section, we discuss limitations and threats to validity.
\vspace*{-.5em}
\subsection{Limitations} 
The experimentation we conducted in this article has a number of limitations that warrant further investigation. The first limitation results from a lack of access to domain experts who could effectively identify genuine cases of incompleteness in the requirements. As a result, we had to resort to simulating incompleteness by withholding content from existing requirements. While this allows for some measure of evaluation for our proposed approach's effectiveness, it may not accurately represent the actual incomplete requirements encountered in real-world scenarios. To develop a more precise understanding of our approach's utility, it is essential to conduct future user studies involving domain experts who can provide insights into the nature and extent of incompleteness in requirements. 

Another potential limitation is the size and diversity of our dataset. Although our dataset covers 15 distinct domains, it may not be representative of all possible scenarios for different industries and contexts. To provide a more comprehensive evaluation of the approach, future experimentation should consider utilizing the entirety of PURE~\cite{ferrari2017pure} and potentially other datasets. 

We used the original BERT model in our experiments due to limited computational resources and were unable to systematically explore the impact of using BERT variants. Although our preliminary experiments did not suggest significant improvements from using BERT variants, further rigorous empirical investigation is needed to confirm this.

Lastly, we note that our research coincided with the release of ChatGPT~\cite{ChatGPT}, which has had a remarkable impact on AI-enabled software engineering research in the past months. While our main proposition, namely using large language models as an external source of knowledge for enhancing requirements completeness, can also be instantiated with ChatGPT and the GPT family of language models~\cite{GPT42023Technical}, we have not yet explored this avenue of research. It is plausible, and in fact likely, that ChatGPT would be a superior alternative to BERT as an external knowledge source for requirements completion and incompleteness mitigation.
\vspace*{-.5em}
\subsection{Internal Validity} 
We intentionally seed incompleteness into requirements to evaluate the effectiveness of our approach in detecting the omissions. We nevertheless recognize that random variation could potentially influence our findings. We employed several  strategies to mitigate the impact of random variation. First, we utilize a substantial dataset consisting of 40 requirements documents, as depicted in Table~\ref{tbl:dataset}. This large dataset helps minimize the effects of random variation, thereby increasing the robustness of our findings. Furthermore, we enhance the reliability of our experiment by repeating each test for a given document five times. This is achieved by shuffling the withheld and disclosed portions of the document, as outlined in Section ~\ref{subsec:proc}. 

{\color{black}%
Our cleaning of the PURE dataset is limited to the simple changes outlined in Section~\ref{subsec:dataset}. We have chosen not to remove non-requirements from these documents. This decision was driven by the lack of clear differentiation between requirements and non-requirements in many PURE documents. Given that the demarcation of requirements can be subjective without domain expertise, we exercised caution and refrained from attempting it. Nevertheless, we  examined the  documents in PURE where an explicit distinction between requirements and non-requirements existed and where the requirements content dominated non-requirements content in terms of token count.
We did not observe subpar results for documents with higher proportions of requirements. In fact, the results for these documents are slightly better than the averages reported. Consequently, we do not believe that the absence of a clear separation between requirements and non-requirements significantly impacted our results. At the same time, when there is an opportunity for an objective separation between requirements and non-requirements, it may be more sensible to scope completeness checking to the requirements statements only.}

\subsection{Construct Validity} We took the following steps to ensure that our metrics are an accurate reflection of the phenomena under investigation. First, we excluded any terminology already present in the disclosed portion of the requirements documents. This process eliminates the possibility of including non-novel terms. Second, we removed duplicates, common words, and stopwords from our analysis. These measures help ensure that we provide an objective assessment of novel terms \hbox{predicted by BERT.}

We further note that we use non-exact matching for assessing the quality of predictions. 
To address potential construct-validity risks associated with non-exact matching, we analyzed the prevalence of non-exact \linebreak matches and also gathered third-party human feedback to obtain unbiased opinions about such matches. As we argued in Section~\ref{sec:discussion}, our estimated error margin for non-exact matching is small and therefore unlikely to significantly impact construct validity.

\vspace*{-.5em}

\subsection{External Validity} 
Our evaluation is based on 40 requirements documents from PURE~\cite{ferrari2017pure}. These documents span 15 different domains and originate from a variety of sources. The size and diversity of our dataset help provide confidence in the generalizability of our findings. Nevertheless, to further enhance external validity, it would be beneficial to conduct experiments with a broader range of requirements documents.

\section{Conclusion}\label{sec:conclusion}
In this article, we explored the usefulness of large language models (LLMs) for detecting incompleteness in natural-language requirements. Specifically, our focus was on evaluating the effectiveness of BERT's masked-word predictions in identifying relevant but absent terminology within requirements.

In lieu of access to human experts and documented instances of incompleteness, we simulated incompleteness  by withholding content from existing documents. Our investigation consisted of three steps. In the first step, we found the optimal number of predictions per masked word, achieving a balance between noise and relevant predictions generated by BERT. In the second step, we evaluated different machine learning classifiers to remove noise from the predictions. Finally, we compared the filtered predictions with terms in withheld section of requirements documents using new, unseen data. Our findings suggest that BERT's masked-word predictions can hint at a sizeable number of instances of incompleteness. Furthermore, the filtering process enhances the accuracy of the predictions by reducing the prevalence of non-relevant terms.

Throughout the research, several areas for improvement came to light. One such area is expanding the dataset used. We limited our analysis to half of the PURE dataset, considering the trade-off between data cleaning effort and computation costs while maintaining a reasonable dataset size. Using the entire PURE dataset, instead of just half, would yield a larger and more diverse sample, and consequently more robust and reliable findings.
Another important area for improvement is our evaluation. Currently, our evaluation involves simulating omissions by withholding data, which provides useful insights but falls short of capturing the practical implications of our approach. {\color{black}An actual evaluation of BERT, used for detecting incompleteness in requirements, will need to develop a practical use case for presenting BERT's predictions to users within the context in which the predictions were derived. Such an evaluation necessarily has to involve domain experts, requirements engineers, and other stakeholders who can vet BERT's predictions, determining whether they represent genuine instances of incompleteness or are superfluous. Our current results merely gauge the potential of BERT for assisting with requirements completeness checking. To determine whether this potential can be transformed into practical benefits, user studies are a must — something that our current work does not perform and is left for future research.}

Finally, noting the fast-evolving landscape of LLMs towards generative models like GPT, it is important to re-evaluate our approach using such models. We anticipate major improvements, particularly with the potential for interactive and conversational exchanges that can incrementally improve requirements completeness.

\sectopic{Acknowledgements.} This work was funded by the Natural Sciences and Engineering Research Council of Canada (NSERC) under the Discovery and Discovery Accelerator programs.

\bibliographystyle{plain}
\bibliography{references}
\clearpage
\appendix
\section*{Appendix}
\section{Psuedo-code for Baselines}\label{sec:pseudo}
\begin{figure}[h]
    \centering
    \begin{verbatim}
Baseline 1:
INITIALIZE novel_common_words to empty list
INITIALIZE all_common_words to list of 250-1000 common English words
FOR word in all_common_words
    IF word is NOT in disclosed_portion
        IF word is NOT in stop_word
                APPEND word to novel_common_words
            END IF
        END IF
    END IF
END FOR
RETURN novel_common_words
\end{verbatim}
    \caption{Baseline 1 Pseudocode}
\end{figure}

\begin{figure}[h]
    \centering
    \begin{verbatim}
Let T be a data frame of terms and their corresponding TF-IDF score
INITIALIZE novel_tfidf_terms to empty list
INITIALIZE tfidf as terms with top k TF-IDF score in T
FOR term in tfidf
    IF term is NOT in disclosed_portion
        IF term is NOT in stop_word
                APPEND term to novel_tfidf_terms
            END IF
        END IF
    END IF
END FOR
RETURN novel_tfidf_terms
\end{verbatim}
    \caption{Baseline 2 Pseudocode}
\end{figure}

\begin{figure}[h]
    \centering
    \begin{verbatim}
INITIALIZE novel_synonyms to empty list
FOR term in disclosed_portion
    INITIALIZE lst_synonyms to synonyms of term
    FOR synonym in lst_synonyms:
        IF synonym is NOT in disclosed_portion
            IF synonym is NOT in stop_word
                    APPEND synonym to novel_synonyms
                END IF
            END IF
        END IF
    END FOR
END FOR
RETURN novel_synonyms
\end{verbatim}
    \caption{Baseline 3 Pseudocode}
\end{figure}
\mbox{\hspace*{1em}}
\end{document}